\documentclass[traditabstract]{aa}
\usepackage[dutch, english]{babel}
\usepackage{amssymb}
\usepackage{amsmath}
\usepackage{mathrsfs}
\usepackage{latexsym}
\usepackage{longtable}
\usepackage{multirow}
\usepackage{epsf}
\usepackage{color}
\usepackage{graphicx}
\usepackage{float}
\usepackage{enumerate}
\usepackage{txfonts}
\usepackage{natbib}
\usepackage{arydshln}
\usepackage{comment}
\usepackage{capt-of}

\newcommand{\one}{~{\sc i}}
\newcommand{\two}{~{\sc ii}}

\newcommand{\mockalph}[1]{}

\begin{document}

\title{Relating jet structure to photometric variability:\\ the Herbig Ae star \object{HD~163296}\thanks{Based on observations performed with X-shooter (program 089.C-0874) mounted on the ESO {\it Very Large Telescope} on Cerro Paranal, Chile}}

\authorrunning{L.E. Ellerbroek et al.}
\titlerunning{Relating jet structure with photometric variability: HD~163296}

\author{L.~E.~Ellerbroek \inst{1}
\and
L.~Podio \inst{2,3}
\and
C.~Dougados \inst{4, 2}
\and
S.~Cabrit\inst{5, 2}
\and
M.~L.~Sitko\inst{6,7, 24}
\and
H.~Sana \inst{8}
\and
L.~Kaper\inst{1}
\and
A.~de~Koter\inst{1,9}
\and
P.~D.~Klaassen\inst{10}
\and
G.~D.~Mulders\inst{11}
\and
I.~Mendigut\'{i}a\inst{12}
\and
C.~A.~Grady\inst{13, 14}
\and
K.~Grankin\inst{15}
\and
H.~van~Winckel\inst{9}
\and
F.~Bacciotti\inst{3}
\and
R.~W.~Russell\inst{16, 24}
\and
D.~K.~Lynch\inst{16, 17, 24}
\and
H.~B.~Hammel\inst{7,18, 24}
\and
L.~C.~Beerman\inst{6, 19, 24}
\and
A.~N.~Day\inst{6, 20, 24}
\and
D.~M.~Huelsman\inst{6,21, 24}
\and
C.~Werren\inst{6, 24}
\and
A.~Henden\inst{22}
\and
J.~Grindlay\inst{23}
}

\institute{Astronomical Institute ``Anton Pannekoek'', University of Amsterdam, Science Park 904, 1098 XH Amsterdam, The Netherlands
\email{l.e.ellerbroek@uva.nl}
\and
Institut de Plan\'{e}tologie et d'Astrophysique de Grenoble, 414, Rue de la Piscine, 38400 St-Martin d'H\`{e}res, France
\and
INAF - Osservatorio Astrofisico di Arcetri, Largo Enrico Fermi 5, 50125, Florence, Italy
\and
CNRS/Universidad de Chile, Laboratoire Franco-Chilien d'Astronomie (LFCA), UMI 3386, Santiago, Chile
\and
LERMA, Observatoire de Paris, UMR 8112 du CNRS, 61 Av. de l'Observatoire, 75014, Paris, France
\and
Department of Physics, University of Cincinnati, Cincinnati OH 45221, USA
\and
Space Science Institute, 4750 Walnut Street, Boulder, CO 80303, USA
\and
Space Telescope Science Institute, 3700 San Martin Drive, Baltimore, MD 21218, USA
\and
Instituut voor Sterrenkunde, KU Leuven, Celestijnenlaan 200B, 3001 Leuven, Belgium
\and
Leiden Observatory, Leiden University, PO Box 9513, 2300 RA Leiden, The Netherlands
\and
Lunar and Planetary Laboratory, The University of Arizona, Tucson, AZ 85721, USA
\and
Department of Physics and Astronomy, Clemson University, Clemson, SC 29634-0978, USA
\and
Eureka Scientific, Inc., Oakland, CA 94602, USA
\and
Exoplanets and Stellar Astrophysics Laboratory, Code 667, Goddard Space Flight Center, Greenbelt, MD 20771, USA
\and
Crimean Astrophysical Observatory, Scientific Research institute, 98409, Crimea, Nauchny, Ukraine
\and
The Aerospace Corporation, Los Angeles, CA 90009, USA
\and
Thule Scientific, Topanga, CA 90290, USA
\and
Associated Universities for Research in Astronomy, Inc., 1212 New York Ave. NW, Washington, DC 20005, USA
\and
Department of Astronomy, University of Washington, Seattle, WA 98105, USA
\and
Department of Physics, Miami University, Oxford, OH 45056, USA
\and
Department of Management Science and Engineering, Stanford University, Stanford, CA 94305, USA
\and
American Association of Variable Star Observers, 49 Bay State Road, Cambridge, MA 02138, USA
\and
Harvard-Smithsonian Center for Astrophysics, 60 Garden Street, Cambridge, MA 02138, USA
\and
Visiting Astronomer, Infrared Telescope Facility, operated by the University of Hawaii under Cooperative Agreement no. NNX-08AE38A with the National Aeronautics and Space Administration, Science Mission Directorate, Planetary Astronomy Program.}

\date{Received; accepted}

%


\date{Received; accepted}

\abstract{
Herbig Ae/Be stars are intermediate-mass pre-main sequence stars surrounded by circumstellar dust disks. Some are observed to produce jets, whose appearance as a sequence of shock fronts (knots) suggests a past episodic outflow variability. This ``jet fossil record'' can be used to reconstruct the outflow history.
We present the first optical to near-infrared (NIR) spectra of the jet from the Herbig Ae star HD~163296, obtained with \textit{VLT}/X-shooter. We accurately determine physical conditions in the knots, as well as their kinematic ``launch epochs". Knots are formed simultaneously on either side of the disk, with a regular interval of $\sim 16$~yr. The velocity dispersion relative to the jet velocity, as well as the energy input, is comparable in both lobes. However, the mass loss rate, velocity, and shock conditions are asymmetric. We find $\dot{M}_{\rm jet}/\dot{M}_{\rm acc}\sim 0.01-0.1$, which is consistent with magneto-centrifugal jet launching models. No evidence for dust is found in the high-velocity jet, suggesting a launch region within the sublimation radius ($<0.5$~au). The jet inclination measured from proper motions and radial velocities confirms it is perpendicular to the disk.
A tentative relation is found between the structure of the jet and the photometric variability of the central source. Episodes of NIR brightening were previously detected and attributed to a dusty disk wind. We report for the first time significant optical fadings lasting from a few days up to a year, coinciding with the NIR brightening epochs. These are likely caused by dust lifted high above the disk plane; this supports the disk wind scenario. The disk wind is launched at a larger radius than the high-velocity atomic jet, although their outflow variability may have a common origin. No significant relation between outflow and accretion variability could be established.
Our findings confirm that this source undergoes periodic ejection events, which may be coupled with dust ejections above the disk plane.
}

   \keywords{Stars: formation -- Stars: circumstellar matter -- Stars: variables: T Tauri, Herbig Ae/Be -- ISM: jets and outflows -- ISM: Herbig-Haro objects -- Stars: individual objects: HD~163296}

\maketitle

\clearpage

\section{Introduction}

Herbig Ae/Be stars (HAeBe) are intermediate-mass ($2-10$~M$_\odot$) pre-main sequence stars. Like their low-mass equivalent, the classical T Tauri stars (CTTS), HAeBe stars are associated with accretion disks and, in a few cases, jets \citep[e.g.][]{Corcoran1998, Grady2000, Grady2004}. Key questions are how exactly these jets are launched and how this process relates to disk accretion \citep{Ferreira2006, Bai2013}. In Herbig systems, the jet-disk coupling may be different than in CTTS. The relatively high stellar luminosity causes grain particles to rapidly evaporate at distances within $\sim 1$~au, creating a dust-free zone \citep{Kama2009}. Accretion is observed in HAeBe stars \citep[e.g.][]{Muzerolle2004, Mendigutia2011}, so an inner gas disk likely exists. Since jets are expected to originate (at least in part) from this region \citep{Blandford1982}, they may help to improve our understanding of the coupling of accretion and outflow in the inner disk.

We may constrain the launching process by observing jet structure and motion. Jets from young stars are usually observed as a sequence of shock fronts or ``knots". These are likely the result of a variable outflow velocity \citep{Rees1978, Raga1990}. A significant asymmetry in velocity and shock conditions is often observed between the two lobes of jets \citep{Hirth1994, Ray2007}. By tracing the trajectories of the knots through space, we are able to reconstruct the epochs when they were formed and their (quasi-)periodic occurence, if any \citep[e.g.,][]{Ellerbroek2013}. On-source photometric and spectroscopic observations that were made during these ``launch epochs'' (when available) may shed light on what happens in the disk whenever a knot is created. These observations may also clarify the origin of the asymmetry between lobes, which is observed in many jet systems. In this paper we combine time-resolved jet and disk observations and diagnostics to constrain the properties of the launch mechanism, the structure of the launch region, the origin of jet asymmetry, and the relation with disk accretion.

The Herbig Ae star HD~163296 is a promising test case for this observing strategy. Its disk is well-studied and associated with a jet, and a copious amount of time-resolved imaging and spectroscopic data are available. Located at a distance of $119\pm11$~pc \citep{VanLeeuwen2007}, the system does not appear to be associated with a star-forming cluster or dark cloud \citep{Finkenzeller1984}. \citet{Meeus2012} do find extended [C\two] emission that may originate from a background or surrounding molecular cloud. 

The bipolar jet HH~409 was discovered on coronagraphic images (and later confirmed with long slit data) of the Space Telescope Imaging Spectrograph (STIS) on the \textit{Hubble Space Telescope} \citep[HST][]{Grady2000, Devine2000}. One of the knots (A) has also been associated with X-ray emission \citep{Swartz2005}. The high-velocity gas in the jet has radial velocities of $200-300$~km~s$^{-1}$. A blue-shifted molecular outflow (up to $13''$ from the source) was found in CO 2--1 and 3--2 emission with the \textit{Atacama Large Millimiter Array} (ALMA) by \citet{Klaassen2013} and also recovered on \textit{Sub Millimeter Array} CO 2--1 images (C. Qi, private communication). The material in the molecular outflow propagates an order of magnitude slower than the fast-moving jet, peaking at $-18$~km~s$^{-1}$ in the systemic rest frame.

The radius of the dust disk around HD~163296 is estimated at around $500$~au, based on the non-detection of the red-shifted jet up to this distance \citep{Grady2000} and the extent of the scattered light emission \citep{Wisniewski2008}. The 850~$\mu$m continuum emission is observed up to $\sim 240$ au; the outer gas disk is detected in CO lines with a Keplerian rotation profile \citep{Rosenfeld2013, deGregorioMonsalvo2013}. 

The near-infrared (NIR) excess has a component that peaks at 3~$\mu$m and is well fitted by a blackbody of $1500$~K. This suggests emission by dust at the evaporation temperature. Interferometric observations \citep{Tannirkulam2008a, Benisty2010a} show that a major fraction of this emission originates from within the theoretical value of the dust sublimation radius ($R_{\rm sub} \sim 0.5$~au for HD~163296, eq.~1 in \citealt{Dullemond2010}). Also, the emission profile is smooth, i.e. does not originate from a sharply contrasted inner disk rim. HD~163296 also shows significant variations in its NIR brightness on timescales of years \citep{Sitko2008}. 

A number of scenarios have been put forward to explain these observations. Hydrostatic disk models are not preferred, as the NIR emission is much stronger than predicted by hydrostatic equilibrium; also, these scenarios do not explain the photometric variability. Alternative scenarios include a dust halo \citep{Vinkovic2006} and a dusty disk wind \citep{Bans2012}. In both cases, dust would not exist within $R_{\rm sub}$, but at larger distances and above the disk plane, making its observation within $R_{\rm sub}$ a projection effect. \citet{Vinkovic2007} suggest that dust ejections, possibly related to jet launching, are responsible for NIR excess and variability. 

The ``fossil record" contained in the jet as described above may help constrain the physical properties and variability of the inner regions of the disk. In this paper we present optical to NIR spectra of the jet and central source. These were obtained with X-shooter on the ESO \textit{Very Large Telescope} (VLT). We combine the spectra and archival images to reveal the outflow history of the system. We then compare this with the photometric variability of the central object. In Sect.~\ref{sec:observations} we describe the newly obtained and archival observational data. In Sect.~\ref{sec:jet} we present our analysis of the jet kinematics and physical conditions. In Sect.~\ref{sec:source}, we present a multi-band lightcurve of the central object compiled from archival and previously unpublished data. We analyze the variability and colors of the source, and estimate its historic accretion rate. In Sect.~\ref{sec:discussion} we discuss the constraints put on the jet launching by our observations. We also propose a possible explanation for the source variability and comment on its relation with jet structure. We present our conclusions in Sect.~\ref{sec:conclusions}.

\begin{table*}[!th]
\begin{center}
\caption{\label{tab:obs}\normalsize{Journal of the X-shooter observations.}}

\begin{tabular}{llllll}
\hline
\hline
Target & HD~163296 & \multicolumn{4}{c}{HH~409}  \\
\hline\\[-8pt]
Date & 6 Jul 2012 & 6 Jul 2012 & 7 Jul 2012 & 2 Jul 2013 & 14 Jul 2013\\
UT (start obs.) & 02:36 & 02:42 & 01:34 & 06:18 & 01:33 \\
HJD - 2400000 (start obs.) & 56114.608 & 56114.613 & 56115.565 & 56475.763 & 56487.565 \\
Slit position angle (N through E) & 42.5$^\circ$ & 42.5$^\circ$ & 42.5$^\circ$ & 42.5$^\circ$ & 42.5$^\circ$ \\
Section covered ($''$) & 0 & $5-25$ & $10-40$ & $2-13$ & $2-13$ \\
(measured from source)  &     & (NE and SW) & (NE) & (NE and SW) & (NE and SW)  \\
Exposure time (s) & 2 & 240 ($5-15''$),  & 300 & 330 & 330  \\
                                 &    & 440 ($15-25''$) &        &          &  \\
Slit width, UVB / VIS / NIR ($\arcsec$) & $0.5~/~0.4~/~0.4$ & $1.0~/~0.9~/~0.6$ & $0.8~/~0.7~/~0.4$ & $1.0~/~0.9~/~0.6$ & $1.0~/~0.9~/~0.6$ \\
Resolution, $\Delta \varv$ (km~s$^{-1}$) & 33~/~17~/~27 & 59~/~34~/~37 & 48~/~27~/~27 & 59~/~34~/~37 & 59~/~34~/~37 \\
$V$-band seeing ($\arcsec$) & $0.8-0.9$ & $0.8-0.9$ & $0.6-0.8$ & $0.6-0.8$ & $0.6-0.8$\\
\hline
\vspace{-1pt}
\end{tabular}
\end{center}
\end{table*}%

\section{Observations, data reduction and archival data}
\label{sec:observations}

In this section we give an overview of the spectroscopic and photometric data presented in this paper. An image of the jet in [S\two] and the definition of the knots can be found in \citet[][W06; their Fig.~1]{Wassell2006}.

\subsection{VLT/X-shooter spectroscopy}
\label{sec:observations:spectroscopy}

Spectra of HD 163296 and its jet were obtained with VLT/X-shooter \citep{Vernet2011}, which covers the optical to NIR spectral region in three separate arms: UVB (290--590~nm), VIS (550--1010~nm), and NIR (1000-2480~nm). Table~\ref{tab:obs} lists the settings and characteristics of the observations. To observe the jet, multiple overlapping pointings of the $11''$ slit were performed. On 6 July 2012, one short on-source exposure was followed by several offsets, covering both lobes of the HH~409 jet up to $25''$. A narrow-slit deeper exposure covering the red lobe up to $40''$ was taken on 7 July 2012. Sky frames were obtained before or after every exposure to correct for telluric emission lines. Since the source is very bright, we have acquired the jet spectra by offsetting the slit with respect to the source. As a result, the inner jet region ($<5''$) is not covered in the 2012 observations. On 2 and 14 July 2013, follow-up observations were carried out in two offsets up to 2$''$ from the source, in order to constrain the positions of the inner knots.

The frames were reduced using the X-shooter pipeline \citep[version 1.5.0,][]{Modigliani2010}, employing the standard steps of data reduction, i.e. bias subtraction, order extraction, flat fielding, wavelength calibration, and sky subtraction, to produce two-dimensional spectra. The wavelength calibration was verified by fitting selected OH lines in the sky spectrum, resulting in a calibration accuracy of a few km~s$^{-1}$. Flux-calibration was performed using spectra of the spectrophotometric standard star GD153 (a DA white dwarf). The width of the knots perpendicular to the jet axis was assumed to be narrower than the seeing, so that the slitlosses were estimated from measuring the seeing FWHM from the spatial profile of a point source on the 2D frame (on the first night, HD163296, on the second night, a telluric standard star). These estimates were refined by comparing the obtained SED to the averaged photometry (see Sect.~\ref{sec:photometry}). They were subsequently corrected for the slit widths used in the jet observations. This procedure results in an uncertainty of about 10\% on the absolute flux calibration; the relative flux calibration is accurate to within 3\%. 

The wavelengths and velocities used throughout this paper are expressed in the systemic rest frame, for which we adopt $\varv_{\rm sys} = 5.8 \pm 0.2$~km~s$^{-1}$  with respect to the Local Standard of Rest, determined by \citet{Qi2011} based on sub-mm emission lines from the outer disk. This value agrees with the observed radial velocity of photospheric absorption lines in the X-shooter spectrum of HD~163296, which give $7.2 \pm 2.0$~km~s$^{-1}$.

\begin{figure*}[!ht]
   \centering
   \includegraphics[width=0.95\textwidth]{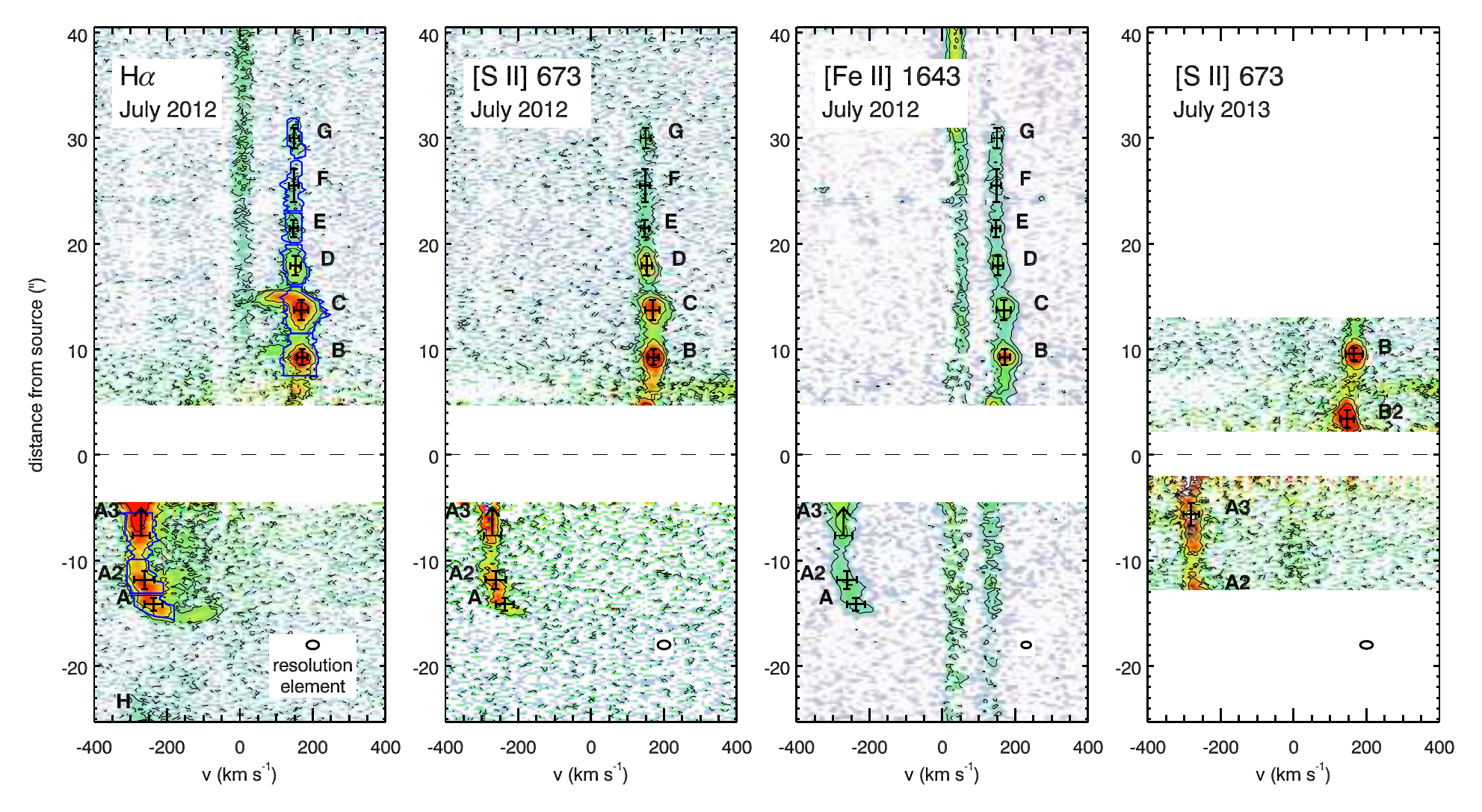} 
   \caption{Position-velocity diagram of H$\alpha$, [S\two]~$\lambda$673~nm, and [Fe\two]~$\lambda$1643~nm in July 2012, and [S\two]~$\lambda$673~nm in July 2013. The $y$-axis denotes the position along the slit; $y>0$ corresponds to the NE lobe, while $y<0$ corresponds to the SW lobe. Due to the brightness of the source, the central region was not observed. Black contours correspond to $\log F_\lambda / [$erg~s$^{-1}$~cm$^{-2}$~$\AA^{-1}$] $= (-17.5,-17,-16.5,-16,-15.5)$. The blue contours indicate the 3$\sigma$ detection level of the knots. The black crosses denote the derived positions and velocities of the knots, and their 1$\sigma$ uncertainties. Note the bow shock features in knot A and C in H$\alpha$.}
   \label{fig:knotdef}
\end{figure*}

After reduction, the frames were merged, averaging the overlapping regions between observations. In the observations closest to the source, a reflection (``ghost spectrum'') of the central source was present at the edges of the slit. We subtracted this contribution in the following way. In a region around every emission line of interest, at every spatial row in the 2D frames, we divided the spectral profile by the on-source spectrum. We then fitted a zero-order polynomial to the residual. Finally, the original spectral profile was divided by the source spectrum multiplied by this fitted value. This results in a position--velocity diagram free of source contamination (Fig.~\ref{fig:knotdef}). Faint, residual continuum emission is seen at $+6-7''$, which likely results from a background continuum source (also visible at this location in the images of \citealt{Grady2000}).

\subsection{Archival data}
\label{sec:observations:photometry}

Measurements of the positions of knots A, B, and C were made on archived HST/STIS imaging and spectroscopy,  HST/Advanced Camera for Surveys (ACS) imaging, and Goddard Fabry-P\'{e}rot (GFP) imaging (\citealt{Devine2000}; \citealt{Grady2000}; W06; \citealt{Guenther2013}, G13). The position of the knots was measured on the [S\two]~673~nm line in the STIS G750L spectra, and on the Ly$\alpha$ line in the STIS G140M spectra.

Multi-wavelength photometric data of HD~163296 were taken from various papers and data catalogs, as well as previously unpublished data. For an overview of these resources, see Table~\ref{tab:archivaldata}. Over the period 1979--2012 the source has been observed with a regularity that varies per band and epoch. The source is best covered in the optical bands. We retrieved data from monitoring campaigns conducted at Madainak Observatory \citep[see also][]{Grankin2007} and at La Silla with the Swiss Telescope in the period 1983--2000. Observations from Las Campanas observatory cover the V-band from 2001--2009 \citep{Pojmanski2004}.

Most of the NIR observations were taken at the Infrared Telescope Facilty (IRTF) and at Mount Lemmon Observing Facility (MLOF) in the period 1996--2009. Part of these data are also presented in \citet{Sitko2008}. Some of the $L$ and $M$ band data were extracted from spectra obtained with The Aerospace CorporationÕs Broad-band Array Spectrograph System (BASS), which covers the 3--13~$\mu$m wavelength region. BASS is described more fully in \citet{Sitko2008}. Additional $JHKLM$ photometric data were obtained using the SpeX spectrograph \citep{Rayner2003}. 0.8--5~$\mu$m  spectra were obtained using a slit width of 0.8$''$ and the echelle diffraction gratings. Zero-point corrections were applied by also observing the star with the prism (0.7-2.4~$\mu$m) and a slit width of 3.0$''$. This method has been shown to provide results comparable to photometric imaging (within 5\%) when the seeing is 1.0 arcsec or better and the airmass does not exceed 2.0 \citep[see][]{Sitko2012}. 

The magnitudes are defined in the Johnson ($UBVRI$), 2MASS ($JHK_{\rm s}$), and ESO ($LM$) filter systems. The measurements have a typical uncertainty of 0.01~mag in the optical and 0.05~mag in the NIR. 



\begin{figure*}[!ht]
   \centering
   \includegraphics[width=0.7\textwidth]{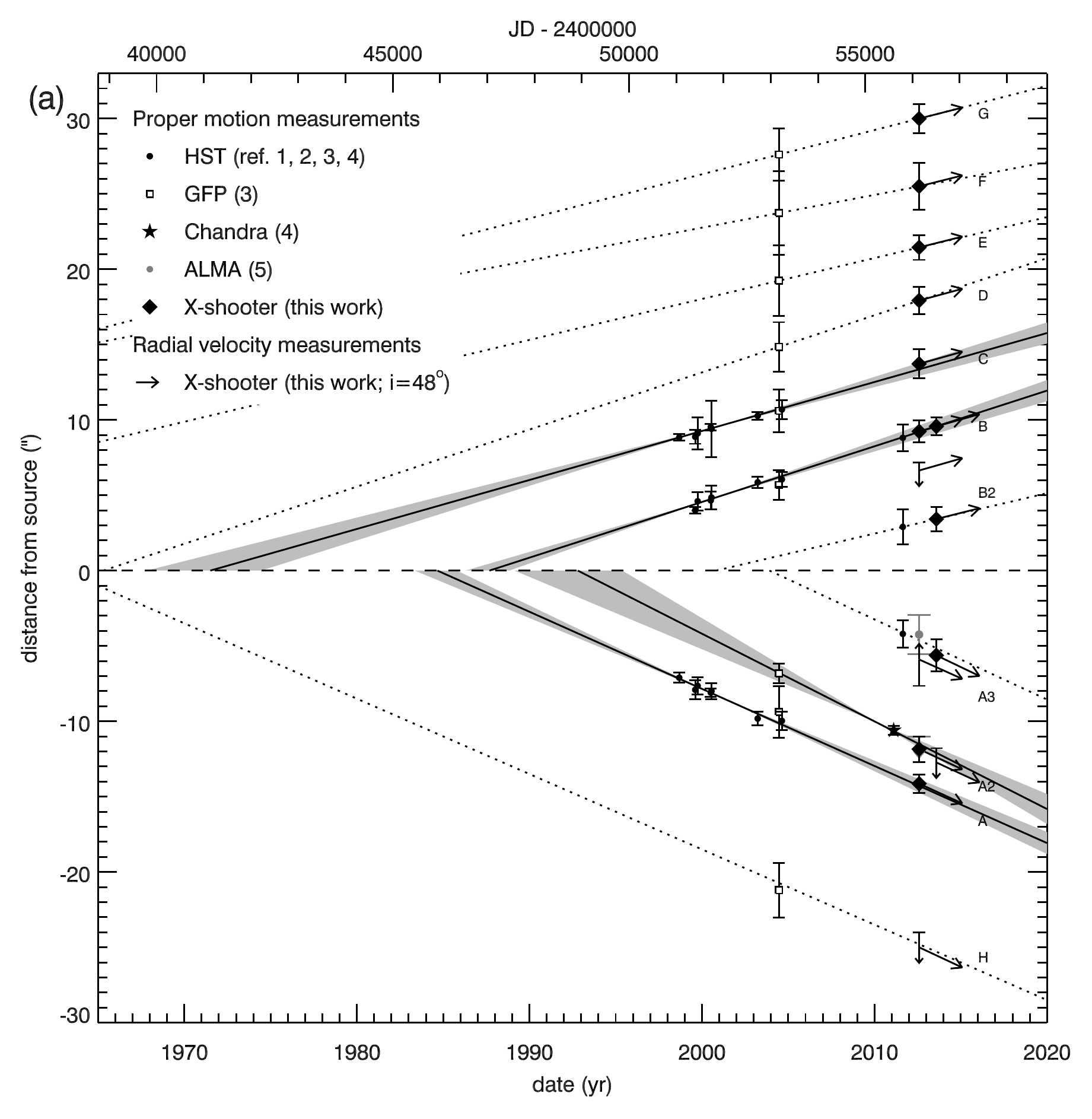} \\
 \includegraphics[width=0.52\textwidth]{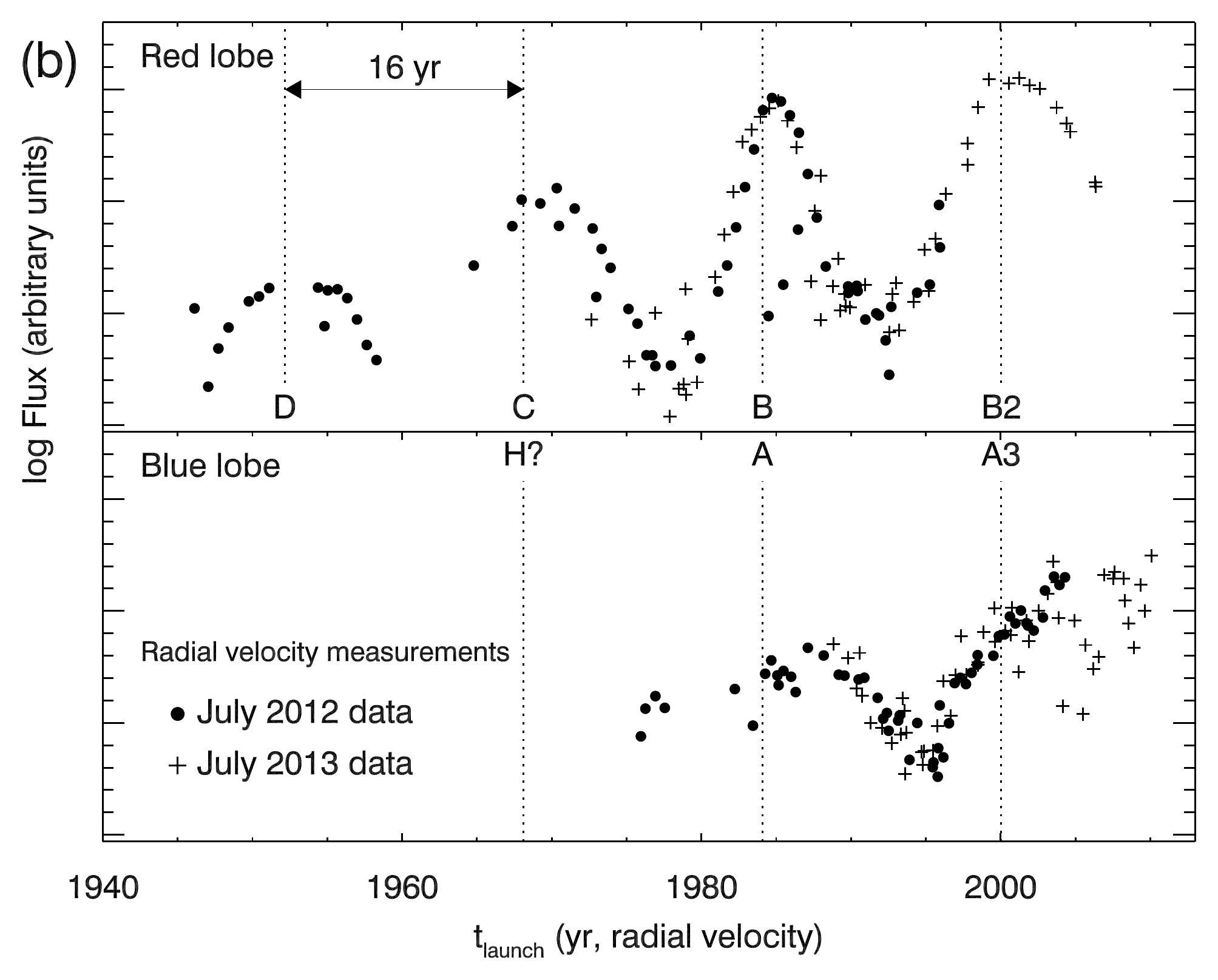}
  \includegraphics[width=0.4\textwidth]{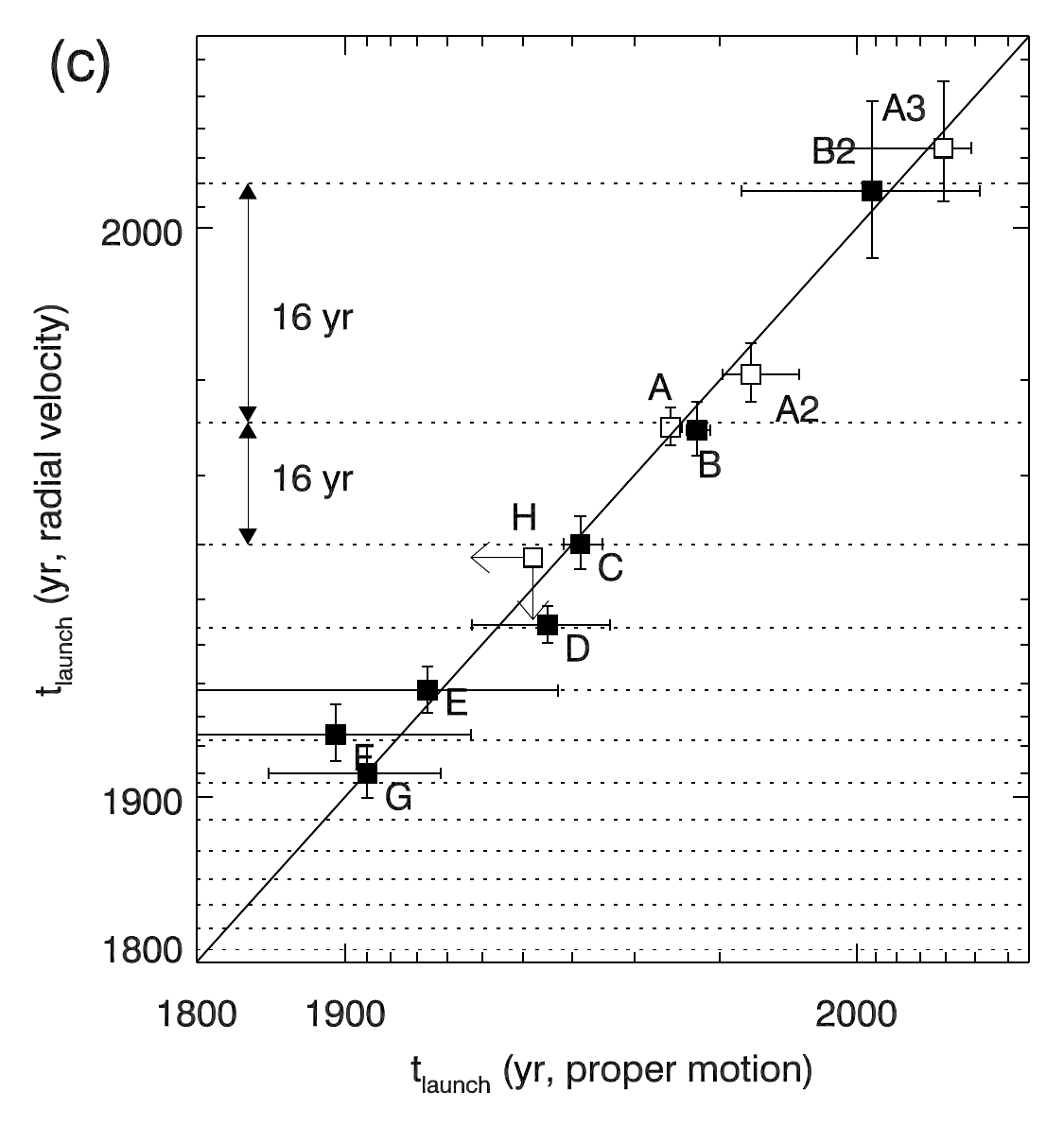}
   \caption{(\textit{a}) Positions of the knots of HH~409 over time; vertical bars correspond to the FWHM of the spatial profile. The shaded bands and dashed lines are linear fits to their trajectories. The arrows represent the velocities on the sky calculated from the radial velocities and disk inclination. References: (1) \citet{Devine2000}; (2) \citet{Grady2000}; (3) \citet{Wassell2006}; (4) \citet{Guenther2013}; (5) \citet{Klaassen2013}.
(\textit{b}) Launch epochs from radial velocities in the X-shooter spectra. Every symbol represents the flux (on a logarithmic scale) and $t_{\rm launch}$ of one row of the two-dimensional spectrum, obtained through the procedure described in the text. Only peaks above the 3$\sigma$ background level are displayed. The data show a periodicity of 16~yr, which was also obtained from the global fit to the jet proper motions (dotted lines). The launch epochs also agree well between the blue and red lobes.
(\textit{c}) Launch epochs of knots in the blue (open squares) and red (closed squares) lobes, derived from proper motions and from radial velocities. The dotted horizontal lines are plotted on a 16-year interval; note the logarithmic scale.
}   \label{fig:propermotions}
\end{figure*}

\section{Results: the jet}
\label{sec:jet}

W06 present an overview of HST data of the jet. It has a position angle of $42.5\pm 3.5^\circ$ (north through east) and extends up to 30$''$ in both the south-west (blue-shifted, or ``blue lobe") and north-east (``red lobe") direction. This section contains an analysis of the X-shooter spectra of the jet. It is divided into two parts: the kinematics and the physical conditions.

\subsection{Kinematics: proper motions vs. radial velocities}
\label{sec:kinematics}

The bipolar HH~409 jet consists of a sequence of local emission line maxima, or ``knots'' (Fig.~\ref{fig:knotdef}). These knots are detected in 40 emission lines of various allowed and forbidden transitions: H\one, [C\one], [N\two], [O\one], [O\two], [S\two], Ca\two, [Ca\two], [Fe\two], and [Ni\two]. No H$_2$ emission is observed along the jet. Fig.~\ref{fig:knotdef} displays the X-shooter spectrum of the H$\alpha$, [S\two]~$\lambda$673~nm, and [Fe\two]~$\lambda$1643~nm emission lines. Knots A, A2, B, C, D, E, F, G, and the onset of H, defined by W06, are indicated in this 2D spectrum. The knot positions are identified by co-adding the position-velocity diagrams of a selection of the strongest emission lines. Two more recently ejected knots: A3 (identified in this work) and B2 (also reported by G13) are only partly covered by the X-shooter slit in July 2012. The movements of these knots are determined from the July 2013 observations and the observations of the Ly$\alpha$ counterpart of knot B2 (G13) and the molecular counterpart of A3 \citep{Klaassen2013}. The jet lobes are asymmetric, the blue lobe being faster ($220-300$~km~s$^{-1}$) than the red lobe ($130-200$~km~s$^{-1}$). Bow-shock features in knot A and C, which are spatially resolved in the HST images (W06), show up as sharp velocity drops in the X-shooter H$\alpha$ position-velocity diagram.

Fig.~\ref{fig:propermotions}a shows the measured positions ($x_{\rm t}$) and lengths (estimated as the FWHM of the spatial profiles along the jet axis) of the knots observed over the 1998\,--\,2013 epoch. The trajectories of knots A, A2, B, and C are well fitted by uniform motion. The trajectories of knots A3, B2, and D--G, though poorly constrained, are similar to the other knots. Assuming constant motion for all knots allows us to make an estimate of their launch epochs ($t_{\rm launch}$). These epochs should then be viewed as the intervals during which the bright knots have formed in the jet, close to the driving source.

Since the knots are regularly spaced, a better constraint on their trajectories is obtained by making a global fit to the data presented in Fig.~\ref{fig:propermotions}a. We assumed that each knot was created simultaneously with its counterpart, that knot creation is periodic, and that knots propagate uniformly on either side of the system. We omitted knot A2 from the fitting procedure, as it has no counterpart. The best-fit model ($\chi^2_{\rm red}=0.82$) has a period $16.0\pm0.7$~yr and proper motions $v_{\rm t, red} = 0.28 \pm 0.01''$~yr$^{-1}$ and $v_{\rm t, blue} = 0.49 \pm 0.01''$~yr$^{-1}$. The global fit is displayed in Fig.~\ref{fig:propermotions_global}.

An independent estimate of the launch epoch estimates is achieved by using X-shooter radial velocities ($\varv_{\rm r}$) and assuming that the jet is perpendicular to the disk equatorial plane. Existing estimates of the disk inclination are based on interferometry of the outer disk ($i_{\rm disk} = 46 \pm 4^\circ$, \citealt{Isella2007}; $i_{\rm disk} = 44^\circ$, \citealt{Rosenfeld2013}) and the inner disk ($i_{\rm disk} = 48 \pm 2^\circ$, \citealt{Tannirkulam2008a}). In this paper we adopt the latter value. From the co-added position-velocity diagrams (Fig.~\ref{fig:knotdef}) of the brightest lines, a spectral profile is extracted at every pixel of width $=0\farcs2$ along the slit. A Gaussian function is fitted to this profile; its centroid velocity is converted to a launch epochs as $t_{\rm launch} = x_{\rm t} / (\varv_{\rm r} \tan i_{\rm disk})$. The fluxes and launch epochs of these individual components are displayed in Fig.~\ref{fig:propermotions}b. The 16-year periodicity in the jet structure, as well as the concurrence of knots in both lobes are evident from this figure.

Fig.~\ref{fig:propermotions}a (arrows) and \ref{fig:propermotions}c illustrate the agreement in the launch epoch estimates between the proper motion and radial velocity methods. The latter figure also demonstrates the periodicity and simultaneity (in both lobes) of the launch events. The jet inclination angle calculated from the average jet velocities is $i_{\rm jet}=\tan^{-1}\langle v_{\rm t}\rangle / \langle v_{\rm r}\rangle = 47\pm2^\circ$. The red-shifted knots B2, B, and C are launched simultaneously with the blue-shifted knots A3, A, and H, respectively. For knot H, $t_{\rm launch}$ is estimated from its position in 2004 and the average $\varv_{\rm r}$ in the blue lobe. Due to their high velocities, the counterparts of the red-shifted knots D, E, F, and G are likely located tens of arcseconds beyond knot H in the blue lobe and are not covered by our observations. The most recent launch epoch traced by our observations peaked in 2001--2004, marking the formation of knots A3 and B2. In the remainder of this paper, we adopt the $t_{\rm launch}$ estimates from radial velocities, as these have the smallest measurement error. 


\begin{table*}[!ht]

\caption{\label{tab:diagnostics}\normalsize{Physical parameters and mass loss rate of HH~409.}}
\begin{minipage}[c]{\textwidth}
    \renewcommand{\footnoterule}{}
        \renewcommand{\arraystretch}{1.6}
\centering
\begin{tabular}{llllllllll}
\hline
\hline
Knot & $x_{\rm t}$ from source & $\varv_{\rm r}$ & $n_{\rm e, post}$ & $X_{\rm e}$ & $\langle n_{\rm H} \rangle$ & $T_{\rm e}$  & $\dot{M}_{\rm jet}$, cr. sect. & $\dot{M}_{\rm jet}$, $L_{\rm [S II]}$ & $\dot{M}_{\rm jet}$, $L_{\rm [O I]}$ \\
 & (July 2012, $''$) & (10$^2$ km~s$^{-1}$) & (10$^2$ cm$^{-3}$) &  & (10$^2$ cm$^{-3}$) & ($10^4$ K) & \multicolumn{3}{c}{($10^{-10}$~M$_\odot$~yr$^{-1}$)} \\
\multicolumn{10}{c}{\textit{Blue lobe}} \\
\hline
A & $ 14.1 \pm   0.6 $ & $  2.4 \pm   0.2 $ & $  4.6 \pm   1.0 $ & $ 0.71 \pm  0.02 $ & $  1.8 \pm   1.0 $ & $  1.3 \pm   0.3 $ & $  2.5 \pm   1.9 $ & $  4.2 \pm   1.2 $ & $  5.1 \pm   1.6 $  \\
A2 & $ 11.8 \pm   0.8 $ & $  2.6 \pm   0.3 $ & $  5.9 \pm   0.8 $ & $ 0.75 \pm  0.01 $ & $  2.2 \pm   1.2 $ & $  1.5 \pm   0.3 $ & $  2.1 \pm   1.7 $ & $  4.2 \pm   0.9 $ & $  5.1 \pm   1.2 $  \\
A3 & $ <  7.6 $ & $  2.7 \pm   0.2 $ & $  6.5 \pm   0.8 $ & $ 0.80 \pm  0.01 $ & $  2.3 \pm   1.3 $ & $  1.5 \pm   0.3 $ & $  1.0 \pm   1.1 $ & $  4.1 \pm   0.9 $ & $  5.0 \pm   1.2 $  \\
\multicolumn{10}{c}{\textit{Red lobe}} \\
\hline
B & $  9.2 \pm   0.7 $ & $  1.7 \pm   0.2 $ & $  6.7 \pm   0.3 $ & $ 0.32 \pm  0.01 $ & $  5.8 \pm   3.2 $ & $  1.0 \pm   0.1 $ & $  2.5 \pm   2.3 $ & $   11 \pm     1 $ & $   13 \pm     2 $  \\
C & $ 13.7 \pm   1.0 $ & $  1.7 \pm   0.2 $ & $  3.4 \pm   0.2 $ & $ 0.43 \pm  0.03 $ & $  2.2 \pm   1.2 $ & $  1.1 \pm   0.1 $ & $  1.9 \pm   1.4 $ & $  8.6 \pm   1.6 $ & $   10 \pm     1 $  \\
D & $ 17.9 \pm   0.9 $ & $  1.5 \pm   0.1 $ & $  2.4 \pm   0.3 $ & $ 0.21 \pm  0.01 $ & $  3.2 \pm   1.8 $ & $  0.9 \pm   0.1 $ & $  4.4 \pm   3.0 $ & $  6.6 \pm   1.3 $ & $  8.0 \pm   1.6 $  \\
\hline
    \vspace{-20pt}
\end{tabular}
\end{minipage}

\end{table*}

\begin{figure}[!ht]
   \centering
   \includegraphics[width=0.85\columnwidth]{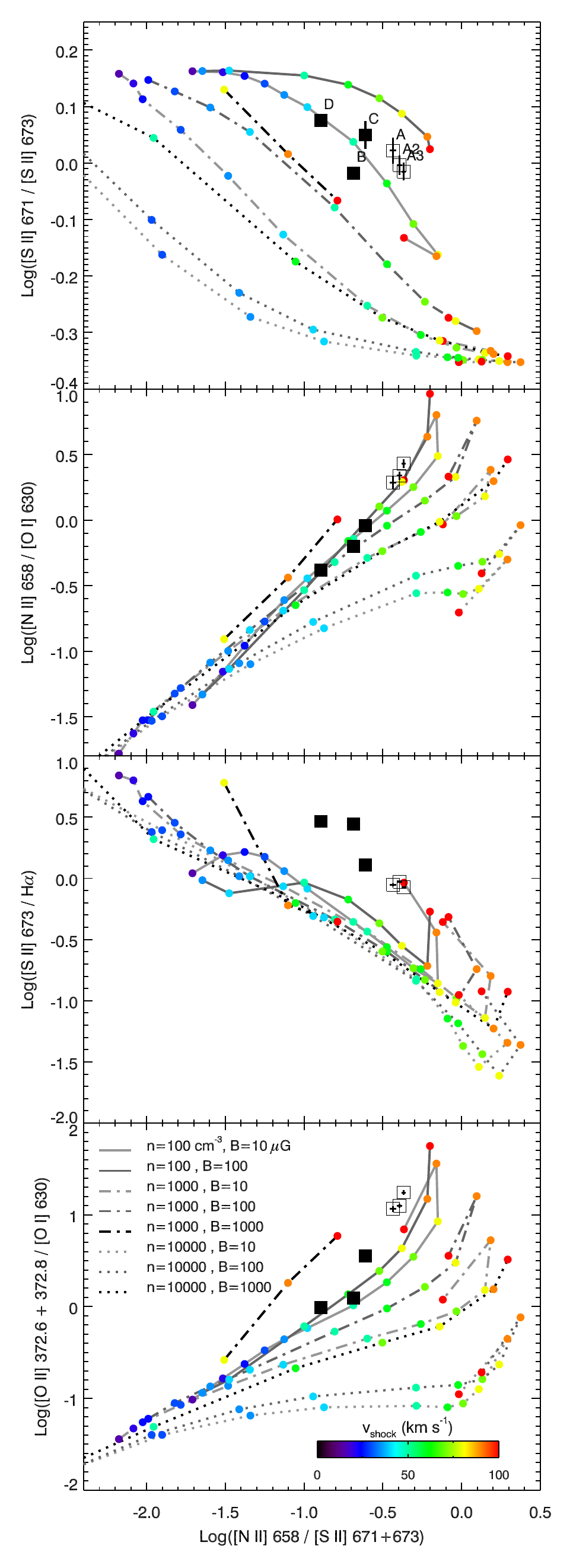} 
   \caption{Emission line ratios for six knots in the blue (open symbols) and red (filled symbols) lobes. Contours indicate the predicted ratios by the models of \citet{Hartigan1994} for a grid of pre-shock densities, magnetic field strengths, and shock velocities. The colored symbols correspond to models with	 $\varv_{\rm shock}=(15,20,25,30,35,40,50,60,70,80,90,100)$~km~s$^{-1}$.}
   \label{fig:ratios}
\end{figure}

\subsection{Physical conditions}
\label{sec:physicalconditions}


The line emission along the jet is a result of collisional excitation in shock fronts. Conditions may vary significantly between the regions up- and downstream of the shock front (``pre"- and ``post"-shock, respectively; \citealt{Hartigan1994}). The physical conditions of the shocked gas in the knots (post-shock electron and total density, $n_{\rm e, post}$, $n_{\rm H, post}$; temperature,  $T_{\rm e}$; and ionization,  $X_{\rm e}$) can be estimated by comparing observed line ratios with those predicted by collisional excitation models (\citealt{Bacciotti1999}; for an application to X-shooter spectra see e.g. \citealt{Ellerbroek2013}). Similarly, the pre-shock parameters (shock velocity, $\varv_{\rm shock}$; pre-shock density $n_{\rm H, pre}$; and magnetic field, $B$) can be obtained by comparing line ratios to shock models \citep[e.g.,][]{Hartigan1994}. 

The results presented in this section are summarized in Table~\ref{tab:diagnostics}. We consider the emission $> 3\sigma$ above the background noise. This is the case for knots A, A2, and A3 in the blue lobe and B, C, and D in the red lobe. The line flux per knot is obtained by integrating over the blue contours seen in Fig.~\ref{fig:knotdef}. We consider only the high-velocity emission. We did not correct for the spatial extent of the knots beyond the slit aperture, which we assume to be low-velocity emission. The line fluxes predicted by models and used in diagnostics in this section are corrected to match the solar abundances of \citet{Asplund2005}, which results in a scaling of typically 0.1~dex.

\subsubsection{Extinction}

Extinction by dust affects the observed line ratios and hence the estimated physical and shock parameters. The value of $A_V$ can be estimated by considering forbidden lines from the same upper level which are far apart in the spectrum. [Fe\two] infrared lines are well suited for this purpose, although large uncertainties in the Einstein coefficients affect the outcome \citep[see][for a discussion]{Nisini2005}. Alternatively H~{\sc i} lines can be used, but their ratios depend also on other model parameters, like the shock velocity \citep{Hartigan1994}. 

We find that within the errors the [Fe\two]~1643/1321~nm and 1643/1256~nm line ratios are consistent with theoretical ones computed for $A_V=0$. Similarly, the H$\alpha$/H$\beta$ line ratios follow the predictions from the \citet{Hartigan1994} shock models for the observed range in $\varv_{\rm shock}$ (see Sect.~\ref{sec:physicalconditions:shockmodels}). Thus, we adopt $A_V=0$ in the knots along the jet. This is consistent (assuming a normal dust-to-gas ratio in the ISM) with the absence of interstellar (gas) features in the spectrum of the central source and the jet \citep{Finkenzeller1984, Devine2000}, as well as the low on-source extinction measured from photometry (see Sect.~\ref{sec:photometry}).

\subsubsection{Electron density, ionization fraction, electron temperature}
\label{sec:physicalconditions:density}

The electron density, hydrogen ionization fraction, and electron temperature of the shocked gas in the knots are calculated from the BE method \citep{Bacciotti1999}. We find an electron density of $\sim 500$~cm$^{-3}$ that decreases along the jet. The ionization fraction is higher in the blue lobe ($0.7-0.8$) than in the red lobe ($0.2-0.4$). The electron temperature decreases away from the source and is higher in the blue lobe ($T_{\rm e} \sim 1-1.5 \times 10^4$~K) than in the red lobe ($T_{\rm e} \sim 10^4$~K). Excitation conditions are thus higher in the blue lobe than in the red lobe. 

These trends are similar to those found by W06 and G13, although the values for $n_{\rm e, post}$ found by these authors are an order of magnitude higher than our calculations. This can be due to the fact that the slit used in these studies (with a width of $0\farcs2$) only samples the central region of the jet, which is likely to have a higher density \citep[see e.g.,][]{Bacciotti2000, Hartigan2007}. The differences in these and other parameters are consistent within the large uncertainties in the line fluxes measured by these authors. The total density in the post-shock region is estimated as $n_{\rm H, post} = n_{\rm e, post}/X_{\rm e}$, under the assumption that hydrogen atoms are the main donor of free electrons. 

\subsubsection{Shock velocity, magnetic field strength, compression}
\label{sec:physicalconditions:shockmodels}

Fig.~\ref{fig:ratios} displays the values of a selection of line ratios against those predicted by \citet{Hartigan1994} for a grid of shock models with $n_{\rm H, pre} = (10^2, 10^3, 10^4)$~cm$^{-3}$, $B=(10, 10^2, 10^3)~\mu$G, and $\varv_{\rm shock}=10-100$~km~s$^{-1}$. The observed ratios are best represented by the models with pre-shock electron densities $n_{\rm H, pre}=100$~cm$^{-3}$, while the magnetic field strength $B$ can adopt values of $10-100~\mu$G. Finally, the observed line ratios indicate that in the blue lobe the shock velocity is higher ($\varv_{\rm shock} \sim 80-100$~km~s$^{-1}$) than in the red lobe ($\varv_{\rm shock} \sim 35-60$~km~s$^{-1}$). This is consistent with the higher excitation conditions in the blue lobe (see also W06 and G13).

The pre-shock density is of order $100$~cm$^{-3}$, as indicated by the comparison with shock models. For the estimated shock velocities {\bf and a typical magnetic field strength of $30$~$\mu$G}, the compression factor, $C = n_{\rm H, post} / n_{\rm H, pre}$ varies between 6 and 20 \citep[Fig. 17 of][]{Hartigan1994}. We adopt $C=13\pm 7$ in all the knots. The average density is estimated as the geometric mean of pre- and post-shock densities. This can be expressed in terms of the post-shock conditions and the compression factor as $\langle n_{\rm H} \rangle = n_{\rm e, post}\, X_{\rm e}^{-1}\, C^{-1/2}$. \citep{Hartigan1994}. The resulting total number density in the jet is of order 100~cm$^{-3}$. The [S\two]/H$\alpha$ ratio in the red lobe is underpredicted by all models except those with $n=1000$~cm$^{-3}$ and $B=1000$~$\mu$G. This may indicate that a strong magnetic field inhibits compression and enhances the emission from lower ionization species.

\subsubsection{Mass loss rate}
\label{sec:physicalconditions:massloss}

The mass loss rate $\dot{M}_{\rm jet}$ is an important parameter in jet physics. It determines the amount of energy and momentum injected in the ISM, and its ratio with the accretion rate reflects the (in)efficiency of the accretion process. Various methods have been developed to calculate $\dot{M}_{\rm jet}$ from the physical parameters \citep[for a review, see][]{Dougados2010}. 
We apply two of these to our dataset: (1) we calculate the mass loss rate from the jet cross section ($\pi R_{\rm J}^2$), total density $\langle n_{\rm H} \rangle$, and velocity ($\varv_{\rm J} = | \, \varv_{\rm r} \, | / \cos i$); and (2) from the [S\two] and [O\one] line luminosities.

\paragraph{Cross section:} This method \citep[``BE" method,][]{Bacciotti1999} calculates the mass loss rate from the average density, velocity, and cross section of the jet:
\begin{equation}
\dot{M}_{\rm jet} = \mu \, m_{\rm H} \, \langle n_{\rm H}\rangle \, \pi R_{\rm J}^2 \, \varv_{\rm J},
\end{equation}
where we adopt $\mu =1.24$ for the mean molecular weight. We take the jet radius $R_{\rm J}$ to be equal to half of the jet FWHM at the selected knot position as measured from resolved HST observations in the [S\two] lines (W06). For the knots considered, $R_{\rm J}$ increases from 40~au up to 100~au away from the source.

\paragraph{Line luminosities:} If the physical conditions are uniform within each knot, the mass loss rate is proportional to the number of emitting atoms in the observed volume \citep{Podio2006, Podio2009}:
\begin{equation}
\dot{M}_{\rm jet} = \mu \, m_{\rm H} \, L_{\rm line} \, \left( h \, \nu \, A_{\rm i} \, f_{\rm i} \, \frac{X^{\rm i}}{X} \, \frac{X}{H} \right)^{-1} \, \frac{\varv_{\rm J} \sin i}{l_{\rm t}},
\end{equation}
where $L_{\rm line}$ is the line luminosity; $\nu$, $A_{\rm i}$, and $f_{\rm i}$ are the frequency, radiative rate, and upper level population fraction for the considered transition, respectively. $X^{\rm i}/X$ is the fraction of atoms of the considered species in the $X^{\rm i}$ ionization state and $X/H$ the species' relative abundance with respect to hydrogen. The upper level population, $f_{\rm i}$, is calculated from the statistical equilibrium equations using the values of $n_{\rm e, post}$ and $T_{\rm e}$ calculated from the line diagnostics. The knot length $l_{\rm t}$ is measured along the slit. This method is applied to two lines: [S\two]~$\lambda$673~nm and [O\one]~$\lambda$630~nm. We assume that all sulphur is singly ionized. To compute the ionization fraction of oxygen, we consider collisional ionization, simple and dielectronic recombination, and charge exchange with H. 

The mass loss rates calculated from the line luminosities (Table~\ref{tab:diagnostics}) agree with each other within the uncertainties. The values from the cross section method are significantly and systematically lower than these. This discrepancy may be caused by an underestimated magnetic field strength and hence, an overestimated compression factor. Additionally, upon comparison with the \citet{Hartigan1994} models (Fig.~\ref{fig:ratioplot}), the observed [N\two]~$\lambda$683~nm/[O\one]~$\lambda$630~nm ratios suggest a lower $X_{\rm e}$ than derived from the BE method, resulting in a higher $\dot{M}_{\rm jet}$. The mass loss rate is constant across each lobe; those found for the red lobe are a factor 2 higher than in the blue lobe. The average mass loss rate is $\langle\dot{M}_{\rm jet}\rangle = 5 \pm 2 \times 10^{-10}  \, {\rm M_\odot yr^{-1}}$.


\begin{figure}[!t]
   \centering
   \includegraphics[width=0.9\columnwidth]{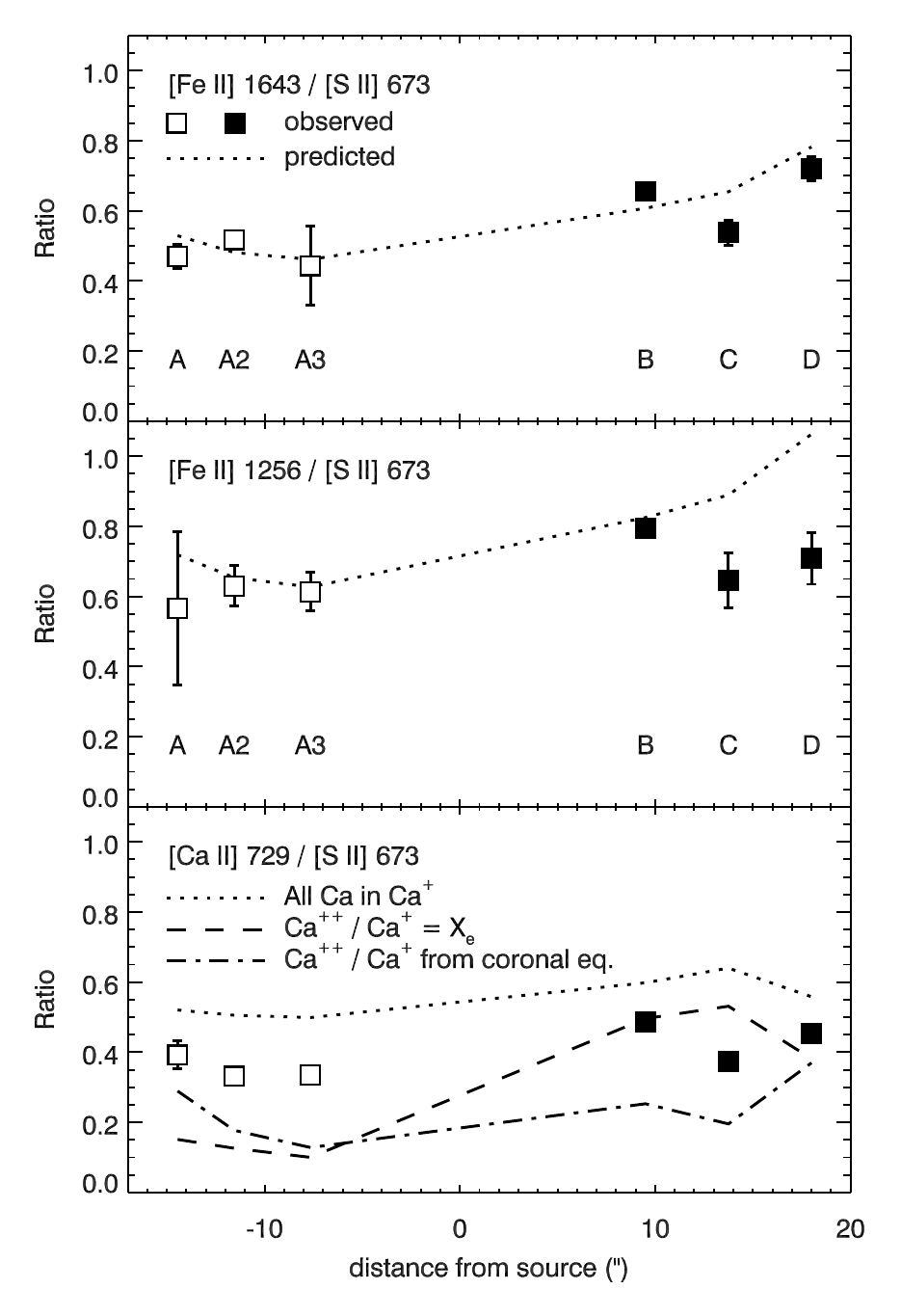}
   \caption{\textit{Top two panels:} Predicted versus observed [Fe\two] / [S\two] ratios, based on the physical conditions derived in Sect.~\ref{sec:physicalconditions}. \textit{Bottom panel:} Same as above, for [Ca\two] / [S\two]. Predictions are made for the three limiting cases discussed in the  text. No evidence is found for depletion of refractory elements in the jet.}
   \label{fig:depletion}
\end{figure}

\subsubsection{Dust content}
\label{sec:physicalconditions:depletion}

The line ratios observed in the jet may also be used to test for depletion of refractory species, such as Ca and Fe. This indicates the presence of dust in the jet launch region. The gas-phase abundance of these species is strongly depleted in the ISM with respect to solar abundances because their atoms are locked onto dust grains \citep{Savage1996}. When dust grains evaporate in the launching region or are sputtered in shocks along the jet, these species are released into the gas-phase \citep{Jones1994, Jones2000, May2000, Draine2004, Guillet2009, Guillet2011}. Gas phase depletion of refractory elements has been observed by \citet{Nisini2005} and \citet{Podio2006, Podio2009, Podio2011} in HH jets from Class I sources at large distances from their driving source. It has also been observed in the inner 100~au of a CTTS jet \citep{AgraAmboage2011}. 

In order to check for the presence of dust grains in the jet launch zone, we estimate the Ca and Fe gas-phase abundance following the procedure illustrated in e.g., \citet{Nisini2005} and \citet{Podio2006}. We compare observed line ratios of refractory (Ca, Fe) and non-refractory (S) species with ratios computed through the estimated parameters ($n_{\rm e, post}$, $X_{\rm e}$, and $T_{\rm e}$). We use the flux ratios of [Ca\two]~$\lambda$729~nm, [Fe\two]~$\lambda$1256~nm, and [Fe\two]~$\lambda$1643~nm to [S\two]~$\lambda$~673~nm (Fig.~\ref{fig:depletion}).

We assume that Fe and S are singly ionized and use the 16-levels model presented in \citet{Nisini2002, Nisini2005} with collisional coefficients by \citet{Nussbaumer1988} for Fe and by \citet{Keenan1996} for S. For the [Ca\two]/[S\two] ratio, it is assumed that no neutral calcium is present, since its ionisation potential is very low, $6.1$~eV. For the ionization balance we consider three limiting cases: 
\begin{itemize}
\item[(1)] All calcium is in Ca$^{+}$. This is likely an overestimate, as some Ca may be doubly ionized. 
\item[(2)] The Ca$^{++}$/Ca$^{+}$ fraction is equal to the hydrogen ionisation fraction, $X_{\rm e}$. As explained in \citet{Podio2009}, this is justified because the ionization potential of Ca$^{+}$ is similar to that of hydrogen (11.9~eV and 13.6~eV, respectively). This also goes for the recombination and collisional ionization coefficients for temperatures lower than $3 \times 10^4$~K. 
\item[(3)] The Ca$^{++}$/Ca$^{+}$ fraction is calculated by assuming coronal equilibrium at the estimated knot temperature $T$. Upward transitions are assumed to be due to electron collisions and downward transitions occur by spontaneous emission. This underestimates the Ca$^{+}$ fraction, as there can be no equilibrium in the jet while the gas is moving.
\end{itemize}

Fig.~\ref{fig:depletion} shows the observed and predicted line ratios. No significant depletion of refractory species is found in the jet knots. Some care should be taken when interpreting this result. The collisional coefficients and atomic parameters for [Fe\two] transitions are affected by large uncertainties \citep{Giannini2008}. The [Fe\two] / [S\two] ratio may be overpredicted if foreground dust is present, which we do not consider likely. The [Fe\two] and [S\two] lines may have different filling factors; the Fe lines could originate from a denser or larger region \citep[see][]{Nisini2005, Podio2006}. When this is taken into account, one should take the estimated Fe gas abundance as an upper limit. Despite these uncertainties, our analysis gives no indication of dust grains in the launch region of the high-velocity jet. 

\begin{figure*}[!ht]
   \centering
   \includegraphics[width=0.8\textwidth, page=1]{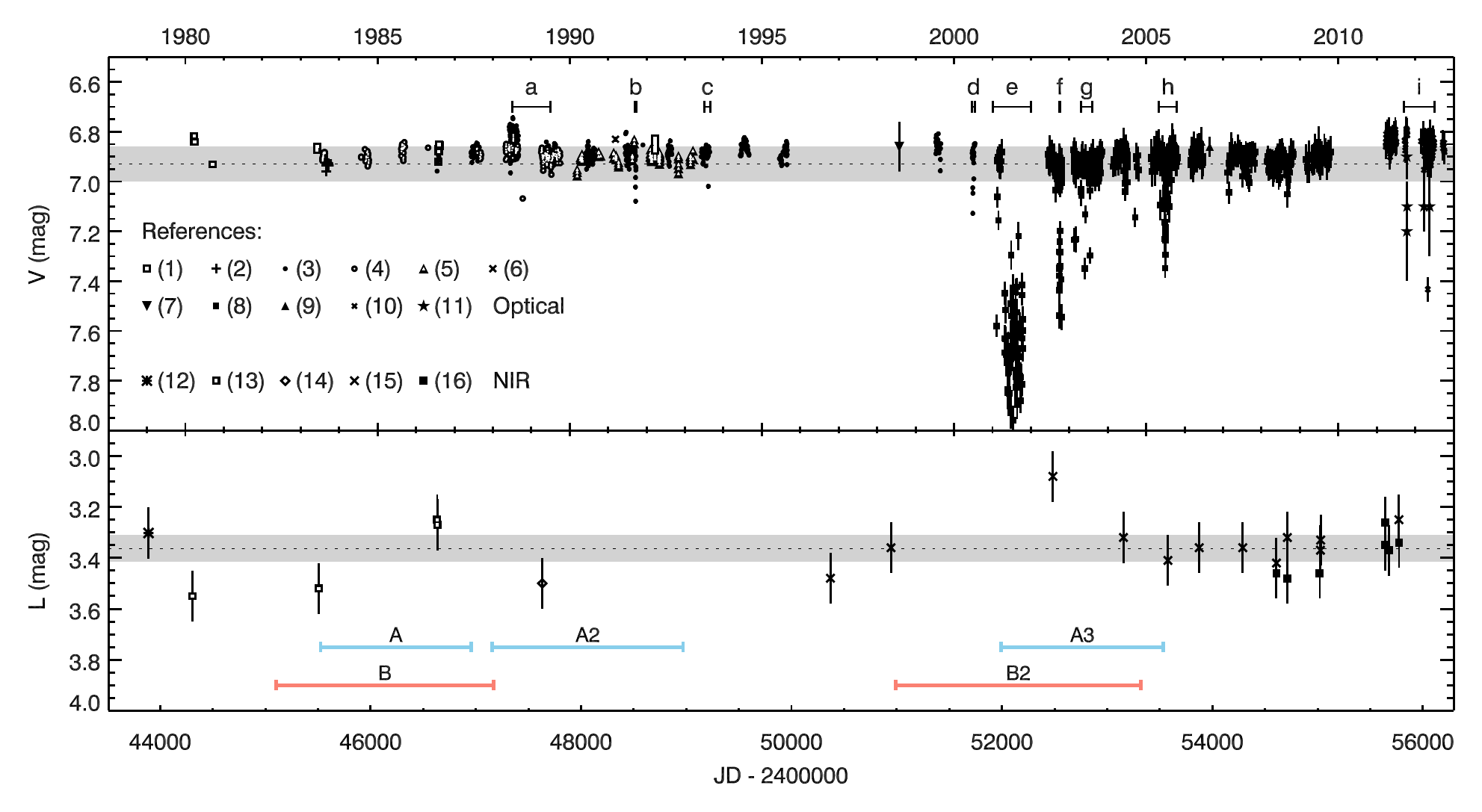}\\
   \includegraphics[width=0.8 \textwidth, page=2]{lightcurve_dips.pdf} 
   \caption{\textit{Top:} Lightcurve of HD~163296 in $V$ (0.55 $\mu$m) and $L$ (3.76 $\mu$m). The dotted line and the gray shaded areas correspond to the mean value and 1$\sigma$ spread. The time intervals denoted above the $V$-band lightcurve are expanded in the bottom panel. Horizontal bars indicate the jet launch epochs estimated from radial velocities (Sect.~\ref{sec:kinematics}). \textit{Bottom:} $V$-band lightcurve for selected intervals, exhibiting the shape, duration, and frequency of the fading events. The calendar date corresponding to the minimum value on the $x$-axis is displayed in the bottom left corner of each graph. References for plot symbols: 
   (1) \citet{DeWinter2001}; 
   (2) \citet{Manfroid1991}; 
(3) Mt. Maidanak (Grankin et al., in prep.); 
(4) Swiss (this work); 
(5) \citet{Perryman1997}; 
(6) \citet{Hillenbrand1992}; 
(7) \citet{Eiroa2001}
(8) \citet{Pojmanski2004}; 
(9) AAVSO (this work); 
(10) \citet{Tannirkulam2008b};
(11) \citet{Mendigutia2013}; 
(12) \citet{Sitko2008};
(13) \citet{DeWinter2001};
(14) \citet{Berrilli1992};
(15) BASS; 
(16) SpeX \citep[][this work]{Sitko2008}. See Table~\ref{tab:archivaldata} for more details.
   }
   \label{fig:lightcurve}
\end{figure*}

\begin{figure}[!ht]
   \centering
   \includegraphics[width=0.95\columnwidth]{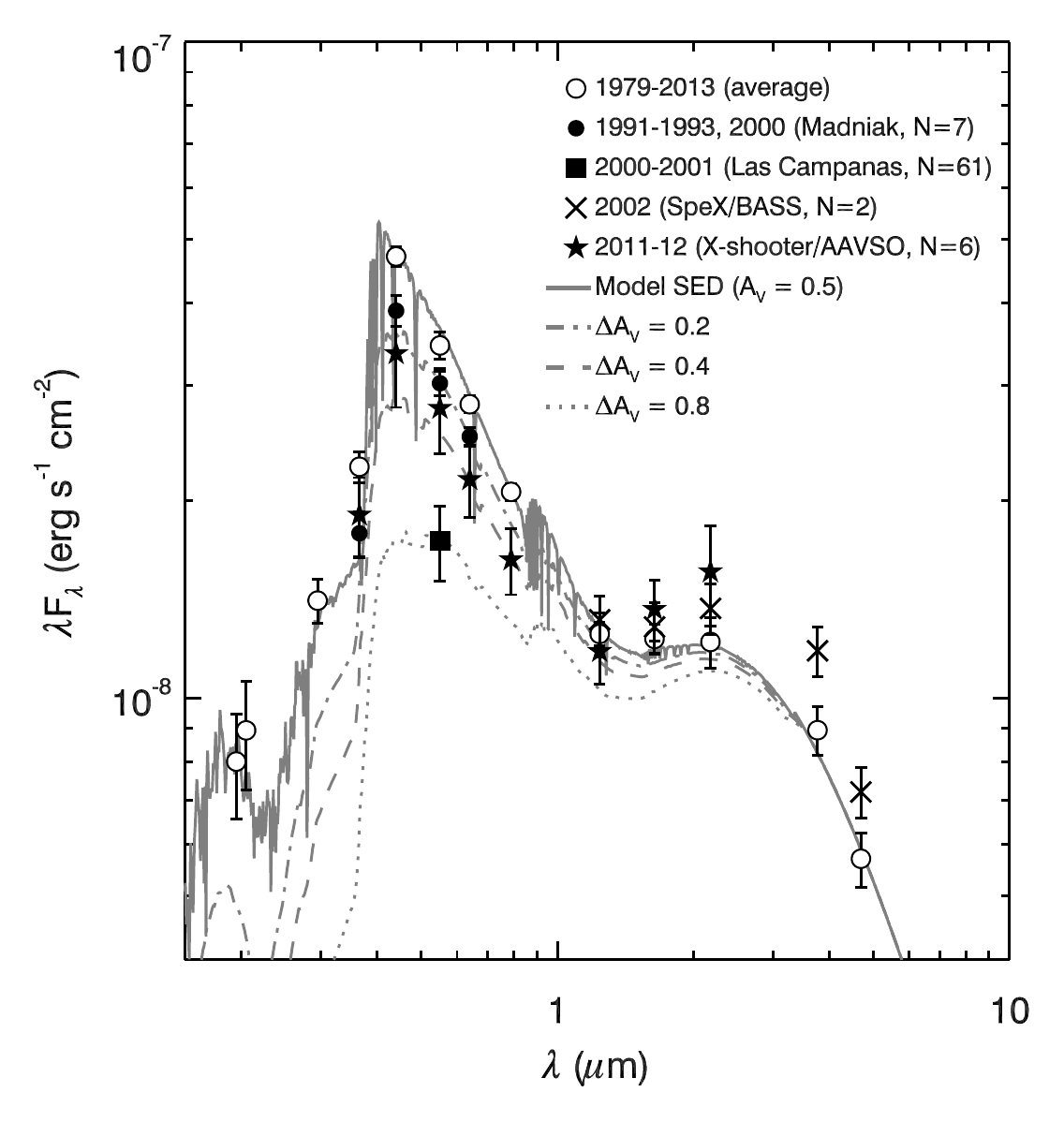} 
   \caption{Average SED of HD~163296 (open symbols), fitted with a 9250~K Kurucz model reddened with $A_V=0.5$; the NIR excess is fitted with a blackbody at 1500~K (the combined model spectrum is shown as a solid gray line). Filled symbols indicate measurements during NIR brightening and optical fading epochs; the number of observations are displayed in brackets. Dash-dotted, dashed, and dotted gray lines represent models reddened with $A_V= 0.7, 0.9$, and 1.3, respectively.}
   \label{fig:sed}
\end{figure}

\begin{figure}[!h]
   \centering
   \includegraphics[width=0.95\columnwidth]{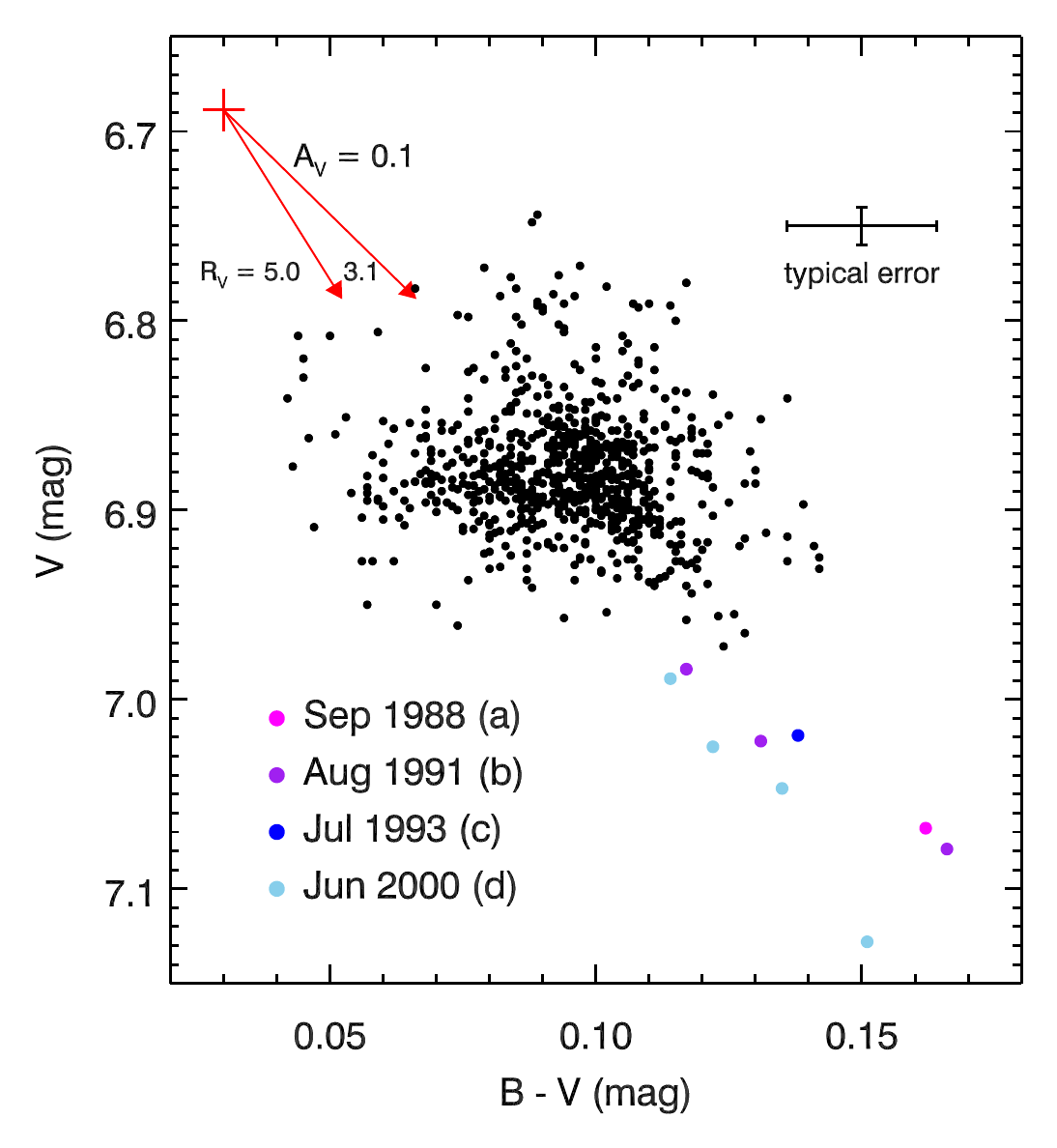} 
   \caption{($B-V,V$) color-magnitude diagram for HD~163296. Datapoints span the period 1983--2012; see Fig.~\ref{fig:lightcurve} for the coverage of this epoch. Only points with errors $<0.1$~mag are plotted. Fading events in 1988, 1991, 1993, and 2000 (letters correspond to Fig.~\ref{fig:lightcurve}) are highlighted with colored symbols. During these events, the colors change along the extinction vector \citep{Cardelli1989}, which is plotted for $R_V = 3.1$ and $R_V=5.0$. The origin of this vector, indicated with a red cross, represents the intrinsic colors and magnitude of an A1V star at 119~pc \citep{Kenyon1995}.}
   \label{fig:colors}
\end{figure}

\section{Results: variability of the central source}
\label{sec:source}

In this section we describe the historic lightcurve of the source, as well as the photometric and spectroscopic accretion diagnostics. We adopt the stellar parameters determined by \citet{Montesinos2009}, which we find to be consistent with the 2012 X-shooter spectrum. These parameters are the effective temperature $T_{\rm eff}=9250 \pm 200$~K , stellar radius $R_*=2.3 \pm 0.2$~R$_\odot$, and mass $M_*=2.2\pm 0.1$~M$_\odot$. 

\subsection{Photometric variability}
\label{sec:photometry}

A lightcurve of HD~163296 is constructed from the collected data described in Sect.~\ref{sec:observations:photometry} and summarized in Table~\ref{tab:archivaldata}. The top panel of Fig.~\ref{fig:lightcurve} displays the $V$- and $L$-band data for the 1978--2013 epoch. Fig.~\ref{fig:lightcurve_all} displays all the photometric data.

The optical brightness of the source over the period 1980--2012 fluctuates around a steady level ($\langle V\rangle=6.93 \pm 0.14$~mag). Over this period, several fading events are seen, during which the optical brightness decreases significantly. The most prominent of these started in March 2001 and lasted for at least 6 months (up to at most 1.1 years). Within one month, $V$ increased up to $0.71 \pm 0.15$~mag above its average value. The bottom panel of Fig.~\ref{fig:lightcurve} displays a ``zoom in" on the lightcurve over this and several other fading events. Fadings of order 0.1 mag occur predominantly in the 2001--2006 epoch with durations from hours to several weeks. 


Optical photometric observations dating back as far as the 1890s were obtained from the DASCH catalog \citep[see][]{Grindlay2012}. Since the target was either saturated or in the non-linear domain on most of the photographic plates, we were only able to retrieve lower limits on the photometric measurements. However, prolonged fadings of more than 3$\sigma$ are observed throughout the lightcurve, indicating that these are recurring events. 

In the NIR, the time coverage is much sparser. The mean brightness in the $L$-band over the 1978--2012 period is $\langle L \rangle = 3.36 \pm 0.10$~mag. The mean brightness level in the $H$, $K$, $L$, and $M$ bands is exceeded by 10\% in three epochs: in 1986 ($N_{\rm obs}=2$, see Fig.~\ref{fig:lightcurve_all}), 2001--2 ($N_{\rm obs}=3$), and 2011--12 ($N_{\rm obs}=4$). The first two epochs were also reported by \citet{Sitko2008}. The 2002 event marked a 30\% increase in flux in the $L$-band. Note, however, that both the measurement error and the intrinsic scatter of the NIR brightness are large. A denser coverage of the lightcurve would be needed to evaluate the significance and uniqueness of such events.


The photometry during fading events is further illustrated in Fig.~\ref{fig:sed}. The average SED was computed by taking the mean of the photometry, omitting measurements that deviate more than 2$\sigma$ from the mean value. It is fitted with a 9250~K Kurucz model and a blackbody at 1500~K. In the literature, estimates for the extinction deviate from $E(B-V)=0.015$ \citep[from the absence of interstellar gas absorption; ][]{Devine2000} up to $E(B-V)=0.15$ \citep[from SED fitting; ][]{Tilling2012}. The average reddening in our photometric data is $E(B-V)=0.065\pm0.016$ (Fig.~\ref{fig:colors}). For the plot in Fig.~\ref{fig:sed} we adopt $E(B-V)=0.15$ and $R_V=3.1$, hence $A_V=0.5$ \citep{Cardelli1989}, which best fits the SED in the $UBVRI$-bands. The disagreement between the extinction estimates may be explained by a circumstellar extinction carrier with a lower gas-to-dust ratio than in the ISM, variability of its column density, or both. The optical colors during the fading events in 1988, 1991, 1993, and 2000 are well fitted by an enhanced dust extinction of $\Delta A_V \sim 0.2-0.4$~mag and $R_V=3.1-5.0$ (Fig.~\ref{fig:colors}). The major fading event in 2001, for which colors are not available, implies an enhancement of $A_V$ by 0.7--0.8~mag. 

The 2002 NIR brightening may have a common origin with the optical fading in 2001. This is further reinforced by the simultaneous measurement of optical fading and NIR brightening in the 2011--2012 X-shooter observations (although large errors affect X-shooter spectro-photometry). We discuss possible explanations for this phenomenon in Sect.~\ref{sec:macc}.






\begin{figure*}[!ht]
   \centering
   \includegraphics[width=0.85\textwidth]{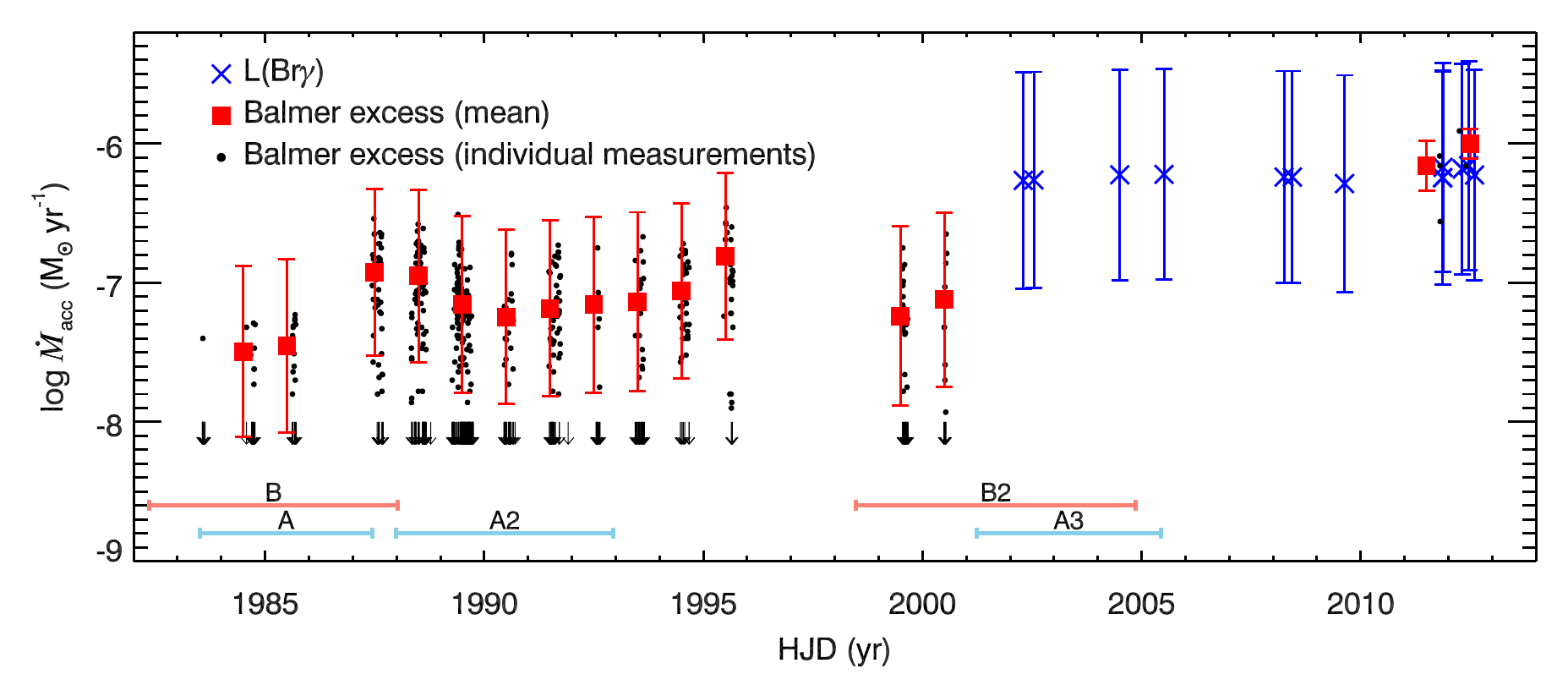} 
   \caption{Accretion rate estimates from the Balmer jump (squares: mean; circles: individual measurements) and Br$\gamma$ luminosity (crosses). Individual measurements through the Balmer jump method have typical errors of 0.5 dex. Jet launch epochs estimated from radial velocities are indicated with horizontal bars (Sect.~\ref{sec:kinematics}).}
   \label{fig:macc}
\end{figure*}

\subsection{Mass accretion rate}
\label{sec:macc}

The mass accretion rate, $\dot{M}_{\rm acc}$, is an important parameter in characterizing disk evolution and the accretion-ejection process. Despite their generally weak stellar magnetic fields, HAeBe stars show signatures indicative of magnetospheric accretion as also observed in CTTS stars. These include UV excess emission from the accretion shock \citep{Mendigutia2011, Donehew2011} and emission line profiles showing infall and outflow signatures \citep{Muzerolle2004}. 

Estimates of $\dot{M}_{\rm acc}$ in HD~163296 using these diagnostics are of order $10^{-8}-10^{-7}$~M$_\odot$~yr$^{-1}$ and vary over more than an order of magnitude \citep{GarciaLopez2006, Eisner2010, Donehew2011, Mendigutia2011, Mendigutia2013}. This wide spread is probably in large part due to the use of different interpretative models in these studies. In this section, we synthesize these different measurements by interpreting them within a uniform method. In this way, any detected variability can more reliably be interpreted as being intrinsic. We review time-resolved measurements of two tracers for $\dot{M}_{\rm acc}$: Balmer excess and Br$\gamma$ emission. 

The most direct diagnostic for the accretion rate is the UV excess emission measured around the Balmer jump. This emission originates in the accretion shock as a release of gravitational energy of the infalling material. The excess emission can be measured by comparing the observed flux to a photospheric model that corresponds to the stellar spectral type and luminosity. This gives an estimate of the accretion luminosity.

Herbig stars are bright, so any UV excess will have a limited contrast with the photospheric emission. Since the photospheric emission is relatively weak blue-ward of the Balmer jump, the $U-B$ color serves as the best diagnostic for the UV excess. The observed $U-B$ color is corrected for extinction by comparing the observed $B-V$ color with the intrinsic color of an A1V star \citep{Kenyon1995}. The Balmer excess is subsequently obtained by comparing the dereddened value of $U-B$ with the intrinsic color. This value is then converted into $\dot{M}_{\rm acc}$ using the accretion shock model described in \citet{Mendigutia2011}. This was done for the datasets of Madainak Observatory, ESO Swiss, and \citet{Mendigutia2013}; see Table~\ref{tab:archivaldata}. 

The second method to estimate $\dot{M}_{\rm acc}$ is based on the Br$\gamma$ line luminosity, $L({\rm Br}\gamma)$. Models predict that the emission lines are formed in the accretion flow \citep{Hartmann1994}. Indeed, a strong correlation is observed between $L({\rm Br}\gamma)$ and accretion luminosity \citep{Muzerolle1998}. This correlation scales up to the intermediate stellar mass regime \citep{Calvet2004, Mendigutia2011}. We use the correlation coefficients from the latter work (their equation 8) to obtain a rough estimate of $\dot{M}_{\rm acc}$.

We derive the Br$\gamma$ line luminosities from the equivalent width ($EW_{\rm obs}$) and $K$ measurements available in the literature. The line luminosity is defined as
\begin{equation}
L_{\rm line} \, = \, 4\pi d^2 \, EW_{\rm cs} \, F_{\rm K} \,10^{0.4 A_K},
\end{equation}
where
\begin{equation}
EW_{\rm cs} = EW_{\rm obs} - EW_{\rm phot} \, 10^{-0.4 \, | \Delta K |}
\end{equation}
is the line equivalent width with respect to the circumstellar continuum. The last term includes a correction factor for the veiling of photospheric lines by the NIR continuum emission. This factor depends on $\Delta K$, defined as the difference between the observed and the photospheric $K$-band magnitude \citep[see also][]{Rodgers2001}; for the latter we adopted $K_{\rm phot}=6.3$~mag. Since in most cases no simultaneous extinction measurement was available, we adopted $A_V=0.5\pm0.5$ for all observations. A photospheric equivalent width $EW_{\rm phot}=-22~\AA$ was assumed \citep{Mendigutia2013}. Table~\ref{tab:lbrg} lists the line luminosities obtained in this way. The corresponding $\dot{M}_{\rm acc}$-values are displayed in Fig.~\ref{fig:macc}.

As fortune would have it, before 2001, just Balmer excess measurements are available, and after 2001 just $L({\rm Br}\gamma)$ measurements, with the exception of the five photometric observations in 2011--12. These were made with X-shooter, thus the two described methods to estimate $\dot{M}_{\rm acc}$ could be simultaneously applied. For a more thorough treatment of the accretion diagnostics of these spectra, we refer to \citet{Mendigutia2013}.

In Fig.~\ref{fig:macc} the measurements from both methods are displayed. The values of $\dot{M}_{\rm acc}$ from the Balmer excess method show no significant peaks, and level around $\log \dot{M}_{\rm acc} = -7.1 \pm 0.7$~M$_\odot$~yr$^{-1}$. Much of the large scatter can be accounted for by the statistical error of $\sim$0.6~dex on the individual measurements. Far-UV spectra were taken in 1986--1987 (with IUE) and 1998--2004 (with HST/FUSE and STIS, see also \citealt{Tilling2012}). The continuum level varies considerably between these measurements. However, the sampling is too sparse to put firm constraints on the accretion rate variability. Within the uncertainties, no significant variation of the accretion rate before 2001 is detected.



The measurements of $L({\rm Br}\gamma)$ in the period 2001--2012 agree well within $2.3\pm0.3\times10^{-3}$~L$_\odot$. The subsequently derived estimates of $\dot{M}_{\rm acc}$ are $\log \dot{M}_{\rm acc} = -6.2 \pm 0.8$~M$_\odot$~yr$^{-1}$. It appears that $\dot{M}_{\rm acc}$ has substantially increased after 2001. The high values measured in 2011--12 agree between the two methods. However, the line luminosity tracer probably suffers from large systematical uncertainties, see Sect.~\ref{sec:discussion:mass}.


\section{Discussion}
\label{sec:discussion}


In this section, we summarize the constraints that our observations of HH~409 put on the jet launching mechanism. We discuss the periodicity and asymmetry of the jet, and the mass loss and accretion rates. Also, we consider the origin of the optical fading events and their possible connection with the accretion and jet launching process. A sketch of the system's proposed geometry is depicted in Fig.~\ref{fig:cartoon}. 

\subsection{Periodicity}

The excellent agreement between radial velocity and proper motion measurements of the individual knots (Sect.~\ref{sec:kinematics}) confirms that each lobe of the jet is perpendicular to the disk within $4.5^\circ$. This is consistent with, but more accurate than previous measurements \citep[][W06]{Grady2000}. 

Over the last decades, bright shock fronts in the high-velocity jet have appeared simultaneously and periodically (period $=16.0 \pm 0.7$~yr) on both sides of the disk. The periodicity of the knot spacings was also noted by W06. The typical error on the launch epochs is a few years; within this uncertainty, they are created simultaneously. Since the sound speed crossing time of the inner disk region ($\sim0.5$~au) is also a few years \citep{Raga2011}, the knots are likely causally connected: a single variation in the outflow mechanism in the inner region may have produced a velocity pulse on both sides of the disk. In Sect.~\ref{sec:discussion:nature} we comment on various explanations for these periodic ``launch epochs".

\subsection{Asymmetry}

The HH~409 lobes are highly asymmetric in velocity and physical conditions. The average velocity in the blue lobe is a factor 1.5 higher than in the red lobe. This asymmetry is reflected in the physical conditions (Sect.~\ref{sec:physicalconditions}): in the blue lobe, the material has higher shock velocities and hence a higher ionization fraction. The mass loss rate in the blue lobe is a factor 2 lower than in the red lobe. Consequently, the energy input in the jet ($\dot{M}_{\rm jet}v^2$) is roughly equal in both lobes. 

The dispersion in the jet velocity $\Delta \varv_{\rm J}$ (defined as the deprojected FWHM of the spectral profile across the slit) is higher in the blue lobe, but the \textit{relative} dispersion $\Delta \varv_{\rm J}/\varv_{\rm J} \sim 0.2$ is similar in both lobes. Interestingly, the same property is observed in other asymmetric jets (RW~Aur, \citealt{Hartigan2009}; DG~Tau~B,  \citealt{Podio2011}). For all three objects, the relative velocity dispersion is higher than what is expected by projection effects given the jet opening angle (see also \citealt{Hartigan2009}). The latter authors suggest the dispersion may be caused by magnetic waves. This would imply that despite the asymmetry in velocity, the Alfv\'{e}nic Mach number $M_{\rm A} = \varv_{\rm J}/\varv_{\rm A}$ is the same in both lobes (and equals $\sim 5$ for HD~163296).

Asymmetry in velocity and ionization conditions is a commonly observed \citep{Hirth1994, Ray2007} and intriguing property of jets. Various explanations have been proposed, including an asymmetric disk structure or magnetic field configuration, or propagation in an asymmetric environment \citep{Ferreira2006, Matsakos2012, Fendt2013}. In some cases, the mass loss rate is found to be similar in both lobes despite the velocity asymmetry, suggesting external conditions cause their different appearance \citep{Melnikov2009, Podio2011}. In other systems, like HH~1042 \citep{Ellerbroek2013}, the mean velocity and mass loss rate is similar on both sides, but velocity modulations are different, indicating a non-synchronized launching mechanism. 

In the case of HD~163296 the similar energy input and simultaneous launching epochs suggest that a single driving mechanism controls jet launching on both sides of the disk. The lobes are asymmetric already close to the source (G13), and no strong density gradients are observed in the surrounding medium \citep{Finkenzeller1984}. The asymmetry may then be the effect of differing conditions in the launching or acceleration regions of each jet lobe.

\subsection{Mass loss and accretion rate}
\label{sec:discussion:mass}

We have measured mass outflow rates of $\dot{M}_{\rm jet} = 0.2-1 \times 10^{-9}$~M$_\odot$~yr$^{-1}$ in the different knots. This is one order of magnitude lower than some of the estimates of W06 and G13. However, the uncertainties in these earlier measurements are quite large because of the low S/N of the spectra used. The absolute value of $\dot{M}_{\rm jet}$ is up to two orders of magnitude lower than sources with similar masses and accretion rates \citep{Melnikov2008, Melnikov2009, AgraAmboage2009, AgraAmboage2011}. In these studies, the mass flux is obtained within a few arcseconds from the source. We measure fluxes further from the source, where outer streamlines which may carry a substantial fraction of the mass. This results in underestimating both the line luminosities and the jet cross section.

The ratio $\dot{M}_{\rm jet}/\dot{M}_{\rm acc}$ is an observable which differentiates theoretical models of jet launching. Measuring it, however, is difficult. Apart from the large systematic uncertainties in the measurements of both $\dot{M}_{\rm jet}$ and $\dot{M}_{\rm acc}$, there is a fundamental issue affecting their comparison. While the accretion rate is measured on source, the mass outflow rate is usually measured in a knot at a distance from the source where it is spatially resolved. The material in this knot has been ejected some time in the past. Moreover, the shocked knot is likely the outcome of an episode of increased accretion activity. Hence, comparing these two measurements probably does not reflect their true ratio at the time of ejection. Such a measurement may be achieved if $\dot{M}_{\rm acc}$ is measured during the creation of a knot. The knot itself should then be observed around one year later, when it can be spatially resolved (but is still reasonably close to the source) to estimate $\dot{M}_{\rm jet}$. 

Our data allow us to make such an estimate. The combined mass loss rate in knots A and B is $\dot{M}_{\rm jet} = 1.3\pm0.6 \times 10^{-9}$~M$_\odot$~yr$^{-1}$. These knots were created in 1983--1988 (Fig.~\ref{fig:propermotions}). During this period the accretion rate averaged $\dot{M}_{\rm acc} = 7.8\pm4.2 \times 10^{-8}$~M$_\odot$~yr$^{-1}$ (Fig.~\ref{fig:macc}). This implies $\dot{M}_{\rm jet}/\dot{M}_{\rm acc}$ is of order $0.01-0.1$, within the range predicted by theoretical models \citep[e.g.][]{Koenigl2000}. The accuracy on this parameter is insufficient to discriminate between different models of jet launching. Averaged over all knots and epochs, $\dot{M}_{\rm jet}/\dot{M}_{\rm acc} \lesssim 0.01$, which may be the result of underestimating the mass loss rate, as described above. 

\begin{figure}[!t]
   \centering
   \includegraphics[width=0.95\columnwidth]{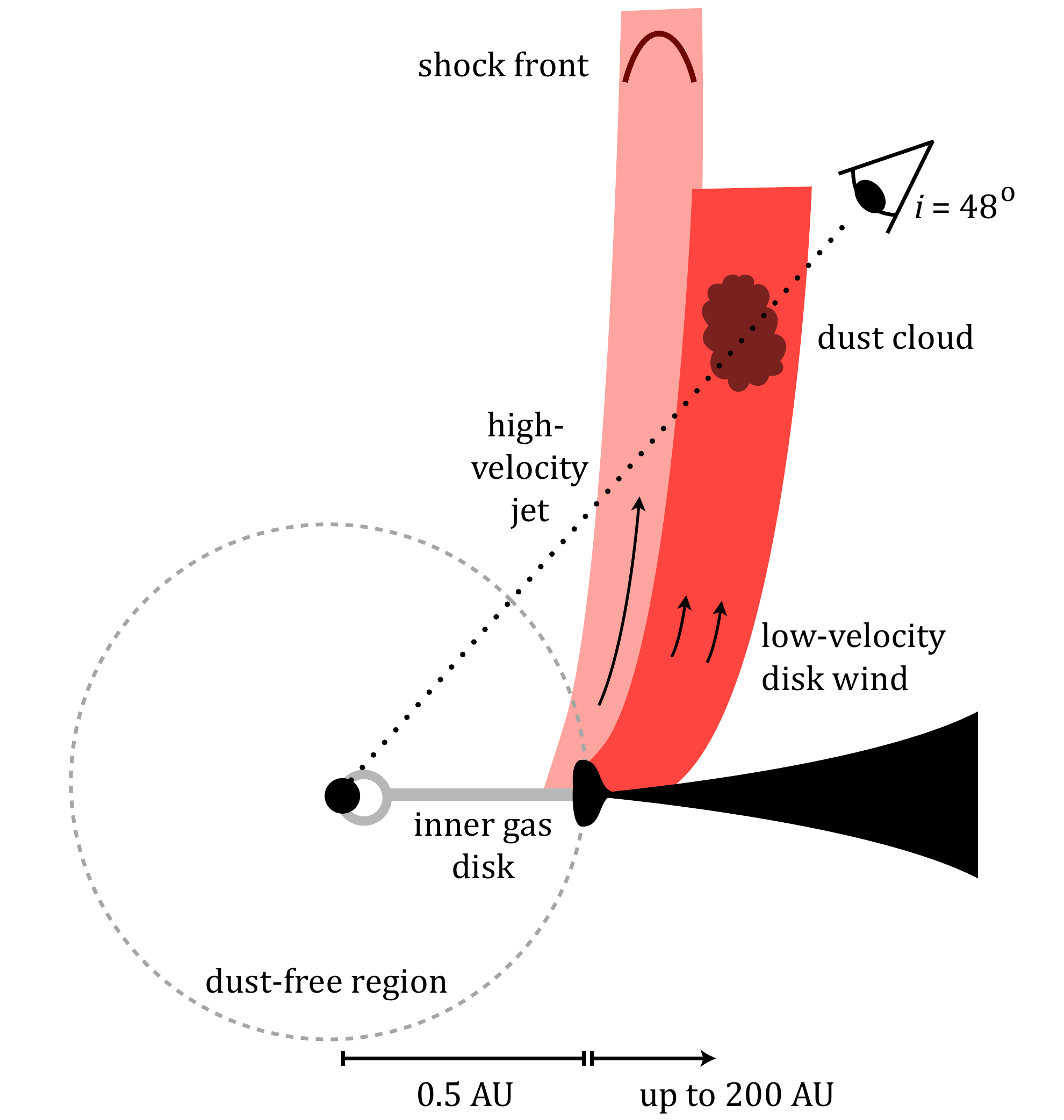} 
   \caption{Cartoon drawing of the geometry of the HD~163296 disk-jet system and a scenario that may explain the optical fading and NIR brightening, as discussed in Sect.~\ref{sec:discussion:dust}.}
   \label{fig:cartoon}
\end{figure}

\subsection{Kinematic structure and size of the launch region}

The jet consists of a layered structure, with a narrow high-velocity component or jet (detected in atomic lines) and a wide low-velocity component or disk wind (detected in molecular lines). The disk wind is wider (several hundred au at $\sim 5''$ from the source, \citealt{Klaassen2013}) than the high-velocity jet (less than 100 au, W06). This ``onion-like" kinematic structure is similar to what has been observed in other jets \citep{Bacciotti2000, Pyo2009, Whelan2010}.

Its low velocity and large radius indicate that the molecular disk wind is launched at larger radii from the central source than the atomic high-velocity jet. This is consistent with predictions from magneto-centrifugal processes, where the velocity in the jet scales with the Keplerian velocity ($\varv_{\rm K}$) at the launch radius as \citep{Blandford1982}:
\begin{equation}
\varv_{\rm J}=\varv_{\rm K}\sqrt{2\lambda_0-3},
\end{equation}
where the magnetic lever arm parameter $\lambda_0=r_{\rm A}^2/r_0^2$ is the squared ratio of the Alfv\`{e}n radius and the launch radius. The results of Sect.~\ref{sec:physicalconditions:depletion} suggest that the high-velocity jet is launched within the dust sublimation radius $<0.5$~au, thus $\varv_{\rm K}\gtrsim60$~km~s$^{-1}$. Given the observed $\varv_{\rm J}$, a lever arm parameter of $\lambda_0 \lesssim 14$ is expected. This is consistent with the estimated $\dot{M}_{\rm jet}/\dot{M}_{\rm acc}$ for a dynamical range in launch radii $r_{\rm 0,max}/r_{\rm 0,min} < 13$ \citep[][their Eq.~17]{Ferreira2006}.

Observational evidence for a disk wind has been found in this system and in other, similar sources. Many spectral lines in HD~163296 exhibit variability related to outflow on timescales of days \citep{Catala1989, Tilling2012}. Interferometric observations of the Br$\gamma$ line show that the emission originates from a region more compact than the continuum, but more extended than the magnetosphere; its most likely origin is a stellar or disk wind \citep{Kraus2008}. Other sources exhibit similar signatures in Br$\gamma$, attributed to an equatorial atomic disk wind within a few au of the star \citep{Malbet2007, Tatulli2007, Eisner2010, Benisty2010b, Weigelt2011}. 

The molecular disk wind is probably launched from outside the sublimation radius and may hence transport dust grains off the disk surface. This is a possible explanation for the fading events, see Sect.~\ref{sec:discussion:dust}.

\subsection{Optical fading, NIR brightening: dust in the wind?}
\label{sec:discussion:dust}

We consider dust clouds entrained in a disk wind as the most likely scenario to explain the observed photometric variability episodes. \citet{Vinkovic2007} propose that this phenomenon may contribute to the infrared variability of young stellar objects. \citet{Bans2012} adapt it to explain the NIR brightening of HD~163296. Our additional observation of optical fading episodes that are well-fitted by dust extinction strengthens this hypothesis. Given the system's inclination and the dust-free region inside 0.5~au, an occulting cloud must be lifted to $\gtrsim 0.5$~au above the disk plane in order to cross the observer's line of sight to the star. The observed diagnostics can be achieved self-consistently with a reasonable set of cloud properties.

The situation is sketched in Fig.~\ref{fig:cartoon}. A dust cloud is lifted from the disk surface and exposed to direct stellar light. As a result, the dust heats up and emits thermal radiation in the NIR. When the cloud transits the star, the latter will appear fainter to the observer. The launch of the cloud in 2001 coincides with the creation of a shock front (knot A3). This knot propagates through the high-velocity jet and is observed far from the source in 2012.

The amount of thermal radiation received at a distance $d$ from a dust cloud of mass $M_{\rm c}$ at a wavelength $\lambda$ in the optically thin limit is: 
\begin{equation}
F_{\lambda, {\rm exc}} = \frac{\kappa_{\lambda} \, B_\lambda(T) \, M_{\rm c}}{d^2},
\label{eq:emission}
\end{equation}
where $\kappa_{\lambda}$ is the opacity of dust, for which we assume a value of $1.34 \times 10^3$~cm$^2$~g$^{-1}$ in the $K$-band, given a typical ISM particle size distribution \citep{Mathis1977}. $B_\lambda(T)$ is the specific intensity of a black-body at the dust temperature, which we assume to be 1500~K. The distance to the observer is $d=120$~pc. To match the observed flux increase of 30\% in the $L$-band in 2002, the cloud mass must be $M_{\rm c} \sim 4 \times 10^{-12}$~M$_\odot$.

Assuming the cloud is co-moving with the CO disk wind at $\sim 10$~km~s$^{-1}$, the observed transit time $t_{\rm tr}=0.6-1.1$~yr implies a size of $1-2$~au. For a spherical cloud we obtain a density $\rho \sim 10^{-18}$~g~cm$^{-3}$. This high value is of the same order as the disk density, suggesting a compact cloud. The optical depth in the $V$-band over a path length $l$ through a medium with density $\rho$ is
\begin{equation}
\tau=\kappa_V \, \rho \, l,
\end{equation}
where $\kappa_V=1.78 \times 10^3$~cm$^2$~g$^{-1}$ for the assumed particle size distribution. This results in an increase of $A_V = \tau/1.086$ with a few tenths of magnitudes during transit. This matches the observed transit depth within an order of magnitude. A more elongated cloud would have to be transported by a wind with higher velocity to match the observed transit time. This would imply a lower density and optical depth than is observed.

An appreciable force is needed to lift the cloud off the disk surface. Based on the estimated $\dot{M}_{\rm wind}$ and observed $v_{\rm wind}$ \citep{Klaassen2013}, the disk wind is easily capable of dragging the cloud along (eq. 24 in \citealt{Kartje1999}, see also \citealt{Bans2012}). The weaker fadings highlighted in Fig.~\ref{fig:lightcurve} may be caused by similar, but smaller clouds lifted off the disk. Alternatively, the dust clouds may be created by condensations in the disk wind \citep[cf.][]{Kenyon1991, Hartmann2004}. 

Additional observations of variability in the 2001--2004 epoch support the dust cloud scenario. \citet{Sitko2008} report an enhancement of the 10~$\mu$m silicate feature in the 2002 observations. \citet{Wisniewski2008} find that the disk is significantly brighter in scattered light emission in 2003--2004 as compared to 1998. \citet{Tannirkulam2008b} report that the $K$-band emitting region in 2003 is larger in size than in 2004--2007. All these signatures are consistent with a variable distribution of dust in the system.

Alternatives to the dust lifting model may be conceived to explain the photometric variability. A non-axisymmetric vertical structure, or warp, in the disk may periodically obscure the central source \citep[cf.][]{Muzerolle2009, Flaherty2012, Bouvier2013}. However, given the inclination, the source of the stellar optical fading has to be located at $h/r \sim 0.9$ (with $h$ the height above the disk surface and $r$ the radial distance from the star). The height of the inner disk of HD~163296 is not expected to exceed $h/r = 0.2$ \citep{Dominik2003}. The typical scale height of inner disk warps is observed to be $h/r \sim 0.3$ \citep{Alencar2010}. This disfavors a disk warp as a likely scenario to explain optical fading. 

Dust may be lifted higher up through tidal disruption of the disk by a companion on a wide and eccentric orbit \citep[cf.][]{Muzerolle2013, Rodriguez2013}. During the periastron passage of the companion the accretion and outflow activity in the disk is expected to be enhanced \citep{Artymowicz1996}. In the next section we comment on this possibility, which may also underly the periodic structure of the high-velocity jet.

\subsubsection{The nature of ``launch epochs"}
\label{sec:discussion:nature}

The creation of knots may be achieved by a variable outflow velocity \citep[e.g.][]{Raga1990, Ellerbroek2013}. We emphasize that the knots should thus not be viewed as ``gas bullets'', but rather as shock fronts propagating through a slower medium. Their periodic creation of knots (during the derived ``launch epochs") does not necessarily have to take place in the disk, but it can be the consequence of (quasi-)periodic processes in the star-disk system. These include disk instabilities \citep[e,g,][]{Zhu2007, DAngelo2012}, the interplay between disk rotation and magnetospheric accretion \citep{Romanova2012}, stellar magnetic cycles \citep{Armitage1995, Donati2003} or interaction with a companion \citep{Muzerolle2013}. The periodicity of 16~yr corresponds to a Keplerian orbit at 6~au. A radial-velocity signal of an unseen stellar or planetary companion at this orbit would at most be a few km~s$^{-1}$. This would have gone undetected in current spectroscopic and (NIR and sub-mm) interferometric campaigns.

Comparable (apparent) periodicities in jet structure have been found in the structure of other HH sources in different evolutionary stages on 10--1000~yr timescales \citep[see][and references therein]{Ellerbroek2013}. This suggests that similar processes that cause jet structure operate in young stellar objects throughout their pre-main sequence evolution. In evolved systems, such as X-ray binaries, long-term variability has been linked with recurring disk instabilities \citep{Dubus2001, Kaur2012}. Oscillations in the gas disks of Be stars produce variability signals with similar periodicity \citep{Okazaki1991, Telting1994}.

It is a compelling possibility that knot creation in the high-velocity jet and dust launching in the (outer) disk wind are the result of the same process. Some of the optical fading and NIR brightening episodes coincide with knot launch epochs (see Fig.~\ref{fig:lightcurve}): event (a) with the launching of knots A and B; (b) with knot A2; and (e)--(h) with knots B2 and A3. This correlation is not well constrained due to the observed spatial sizes of the knots. In addition, the limited coverage of the photometric observations inhibits us to establish a one-to-one relation between photometric fading/brightening and launch epochs. Also, the ``launch period" of $16.0\pm0.7$~yr derived from the spacing and velocities of the knots is not seen in the lightcurve. This prevents us from establishing that one and the same periodic mechanism is responsible for all the jet and photometric variability signatures. 

Also, basic jet launching theory predicts a coupling between accretion and ejection activity \citep{Koenigl2000}. The kinematic structure of jets may then contain a fossilized record of a variable (FU Ori / EX Ori--like) accretion process \citep{Reipurth1997, CarattioGaratti2013}. In HD~163296, a correlation between accretion variability and the jet launch epochs is not apparent given the poorly constrained estimates of $\dot{M}_{\rm acc}$ (Fig.~\ref{fig:macc}). As was mentioned above, \citet{Kraus2008} find that a stellar or disk wind likely contributes significantly to the Br$\gamma$ profile. This may result in a systematical overestimate of the accretion rate from $L({\rm Br}\gamma)$. The apparent increase of the accretion rate after 2001 (Fig.~\ref{fig:macc}) may be strongly biased by this.

\section{Summary and Conclusions}
\label{sec:conclusions}

In this paper, we have presented a kinematical and physical analysis of the HH~409 jet, and have related this to the historic lightcurve of its driving source HD~163296. The main conclusions of this work are listed below. 

\begin{itemize}
\item In the HD 163296 disk-jet system, periods of intensified outflow activity in the high-velocity jet occur on a regular interval of $16.0\pm0.7$~yr.
\item Although physical conditions and mass loss rates are asymmetric between the two lobes, the energy input is similar and the periodic creation of knots is synchronized. The velocity dispersion scales with the jet velocity in both lobes, regardless of their asymmetry. These observations suggests that the origin of the asymmetry resides close to, but not inside the disk.
\item We observe no evidence for depletion of refractory species with respect to their solar abundances -- a proxy for dust formation -- in the high-velocity jet. The absence of dust in the high-velocity jet is consistent with a launch region within $0.5$~au. 
\item Transient optical fading is identified for the first time in this source. This, together with enhanced NIR excess, is consistent with a scenario where dust clouds are launched above the disk plane. 
\item Dust ejections coincide with high-velocity atomic jet launch events (although no one-to-one relation can be established). This tentatively suggests a common origin despite their different launch points (e.g., disk instabilities induced by tidal perturbations). 
\item A direct correlation between accretion and ejection events could not be established given the large uncertainties in $\dot{M}_{\rm acc}$. We do derive $\dot{M}_{\rm jet}/\dot{M}_{\rm acc} \sim 0.01-0.1$, which is consistent with magneto-centrifugal models for jet launching.
\end{itemize}

Jets are rare in the more evolved HAeBe stars; therefore HD~163296 is an excellent example to study the accretion-outflow phenomenon as it scales up to higher masses. The ``jet fossil record" allows us to probe outflow variability on unprecedented timescales. Combining this with photometric variability monitoring also facilitates a direct comparison of accretion and outflow properties, leading to a better understanding of the complex jet launching process. This observing strategy is applicable across a very broad range of astrophysical disk-jet systems that show long-term variability, including young stellar objects, X-ray binaries and active galactic nuclei.


\acknowledgements
The referee, Dr. Tom Ray, is acknowledged for useful comments that helped improve the manuscript. The authors thank Myriam Benisty, Jerome Bouvier, Carsten Dominik, Patrick Hartigan, Henny Lamers, Koen Maaskant, Michiel Min, Brunella Nisini, Charlie Qi, Alex Raga, and Rens Waters for discussions about this work. Patrick Hartigan is also acknowledged for kindly providing the shock model results in tabular form. The ESO staff and Christophe Martayan in particular are acknowledged for their careful support of the VLT/X-shooter observations. Daryl Kim is acknowledged for his technical support of the BASS observing runs. The authors thank Bill Vacca, Mike Cushing, and John Rayner for many useful discussions on the use of the SpeX instrument and the Spextool processing package. This work was supported by a grant from the Netherlands Research School for Astronomy (NOVA), NASA ADP grants NNH06CC28C and NNX09AC73G, and the IR\&D program at The Aerospace Corporation. LP acknowledges the funding from the FP7 Intra-European Marie Curie Fellowship (PIEF-GA-2009-253896). 


\clearpage

\appendix

\section{Additional materials}

\noindent\begin{minipage}{\textwidth}
   \centering
   \includegraphics[width=0.95\textwidth]{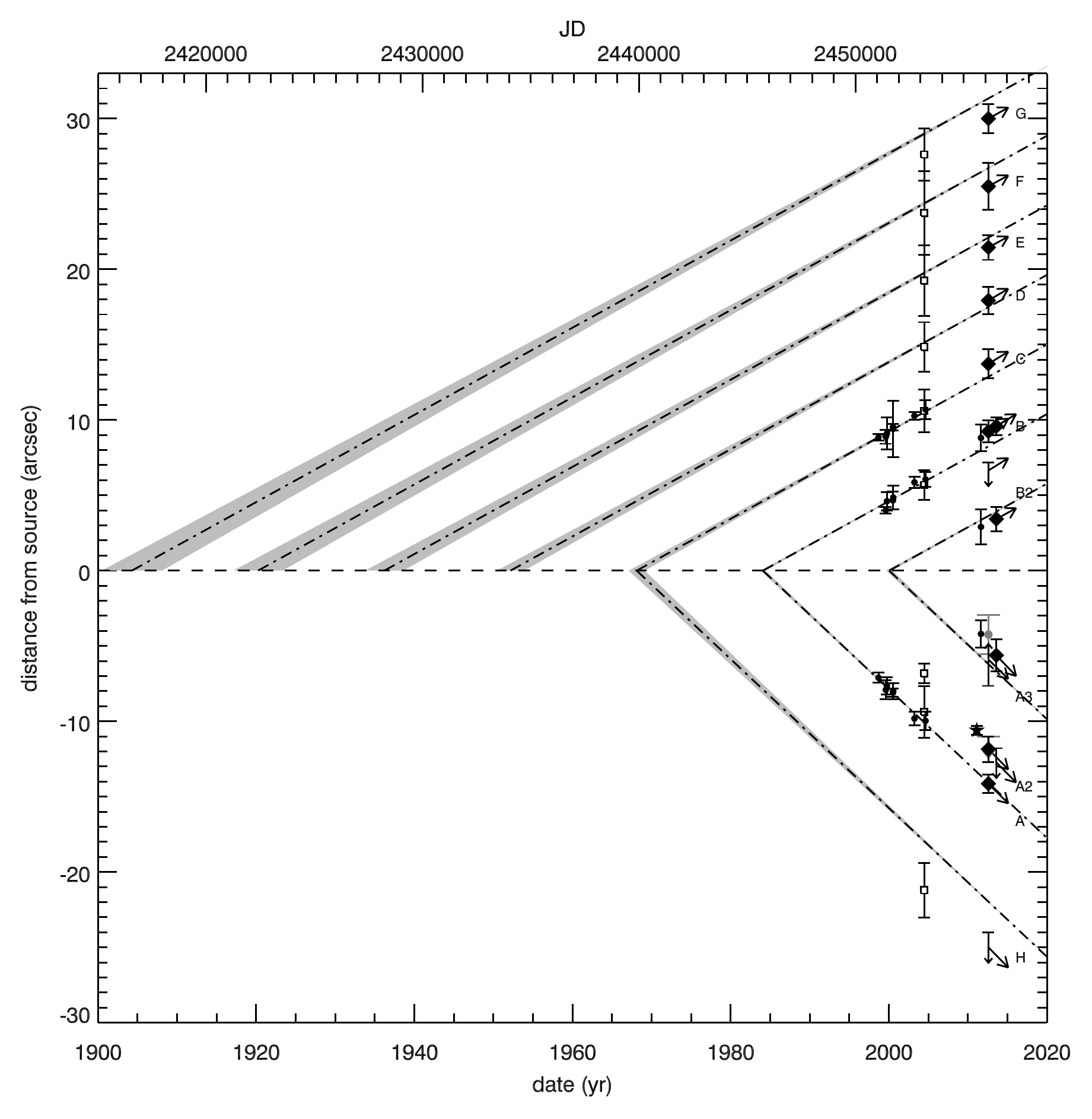} 
   \captionof{figure}{Proper motions of the knots in the HH~409 jet. Same as Fig.~\ref{fig:propermotions}; the dash-dotted lines correspond to the best global fit made to the positions of the knots, with parameters $16.0\pm0.7$~yr and proper motions $v_{\rm t, red} = 0.28 \pm 0.01''$~yr$^{-1}$,  and $v_{\rm t, blue} = 0.49 \pm 0.01''$~yr$^{-1}$. The A2 data are omitted from the fitting procedure. 
   }   \label{fig:propermotions_global}
\end{minipage}
\clearpage

\begin{figure*}[!ht]
   \centering
   \includegraphics[width=0.95\textwidth]{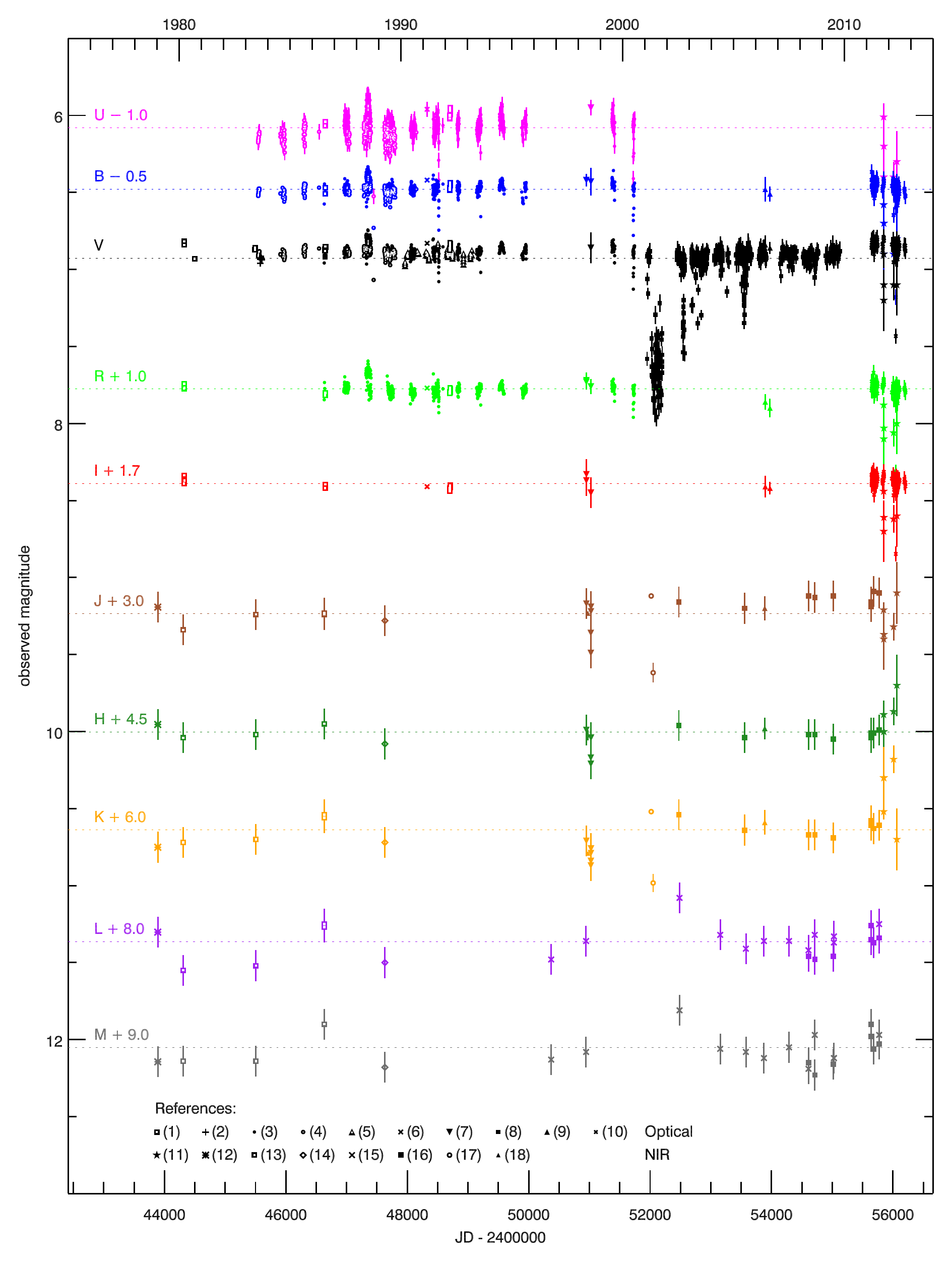} 
   \caption{Lightcurve of HD~163296 compiled from the resources listed in Tab.~\ref{tab:archivaldata}. The dotted lines indicate the mean values. References for plot symbols: 
   (1) \citet{DeWinter2001}; 
   (2) \citet{Manfroid1991}; 
(3) Mt. Maidanak (Grankin et al., in prep.); 
(4) Swiss (this work); 
(5) \citet{Perryman1997}; 
(6) \citet{Hillenbrand1992}; 
(7) \citet{Eiroa2001}
(8) \citet{Pojmanski2004}; 
(9) AAVSO (this work); 
(10) \citet{Tannirkulam2008b};
(11) \citet{Mendigutia2013}; 
(12) \citet{Sitko2008};
(13) \citet{DeWinter2001};
(14) \citet{Berrilli1992};
(15) BASS; 
(16) SpeX \citep[][this work]{Sitko2008}; 
(17) \citet{Kimeswenger2004};
(18) \citet{Skrutskie2006}.
   }
   \label{fig:lightcurve_all}
\end{figure*}

%

\begin{table*}[t]
\begin{center}
\caption{\label{tab:archivaldata}\normalsize{Overview of datasets used in this study.}}
\renewcommand{\arraystretch}{1.5}

\begin{tabular}{llllll}
\hline
\hline
Epoch (HJD) & Facility & Mirror / Instrument & Type\footnote{S: spectra, P: photometry, I: imaging, II: interferometric imaging.} & Wavelength / bands & Reference \\
\hline\\[-8pt]
\multicolumn{6}{c}{\textit{Observations of the jet}}\\
\hline
1998--2011 & HST & STIS  & S, I & FUV, $R$ & \citet{Devine2000};  \\
 & & & & & \citet{Grady2000}; \\
 & & & & & \citet{Wassell2006}; \\
 & & & & & \citet{Guenther2013} \\
2003--2004 & HST & ACS  & I &  $R$ & \citet{Wassell2006}  \\
2004  & Apache Point Obs. & GFP & I &  $R$  & \citet{Wassell2006} \\
2012 & ALMA &  & II & 0.8--1.42~mm &  \citet{Klaassen2013}  \\
2012 & VLT & X-shooter & S & 290--2480~nm &  This work \\
\hline
\multicolumn{6}{c}{\textit{Observations of the source}}\\
\hline
1889--1989 & Various & Various & P & $B$ & This work; \\
 & & & & & DASCH \citep{Grindlay2012}  \\
1979 & KPNO & 1.3 m, Bolo/Otto & P & $JHKLM$ & \citet{Sitko2008} \\
1980--1986 & La Silla & Dutch 90~cm & P & $UBVRI$ &  \citet{DeWinter2001}  \\
1980--1992 & La Silla & ESO 50~cm / 1~m & P & $UBVRI$ &  \citet{DeWinter2001}  \\
1980--1986 & La Silla & ESO 1~m  & P & $JHKLM$ &  \citet{DeWinter2001}  \\
 1983 & La Silla & ESO 50~cm & P & $V$ &  \citet{Manfroid1991} \\
1983--1990 & La Silla & Swiss 1.3 m & P & $UBV$ & This work \\
 1986--2000 & Mt. Maidanak Obs. & 48 cm / 60 cm  & P & $UBVR$ & This work;  \\
 &  &  &  &  &  see also \citet{Grankin2007}  \\
1989 & La Silla & ESO 1~m & P & $JHKLM$ & \citet{Berrilli1992} \\
1990--1993 & Hipparcos &  & P & $V$ & \citet{Perryman1997}     \\
1991 & USNO Flagstaff 1 m & EMI 9658 R & P & $UBVRI$ & \citet{Hillenbrand1992} \\
1996--2009 & IRTF / MLOF 1.5 m & BASS & S & $LM$ & This work; \citet{Sitko2008} \\
1996 &  ISO & SWS & S & $LM$ & \citet{VanDenAncker2000} \\
1998 & Cerro Tololo  & 1.3m & P & $JHK$ & 2MASS \citep{Skrutskie2006} \\
1998 &  Nordic Optical Telescope & Turpol & P & $UBVRI$ &  \citet{Oudmaijer2001}  \\
1998 &  Teide Obs. & Carlos S\'{a}nchez 1.5 m & P & $JHK$ &  \citet{Eiroa2001}  \\
2001 & La Silla & ESO 1m, DENIS & P & $JK$ & \citet{Kimeswenger2004} \\
2001--2009  & Las Campanas Obs.  & ASAS & P & $V$ & \citet{Pojmanski2004}\\
2002--2005 & Lick 3m & NIRIS / VNIRIS & S & $VRJHKLM$ & \citet{Sitko2008} \\
2006 &  Kitt Peak & MDM 2.4 m & P & $BVRIJHK$ &  \citet{Tannirkulam2008a}  \\
2006--2011 & IRTF & SpeX & S & $JHKLM$ & This work \\ 
2011--2012 & AAVSO & Bright Star Network & P & $BVRI$ & This work \\
2011--2012 & VLT & X-shooter & S & $UBVRIJHK$ &  \citet{Mendigutia2013}  \\
2012 & VLT & X-shooter & S & $UBVRIJHK$ &  This work  \\
\hline
\vspace{-17pt}
\end{tabular}
\end{center}
\end{table*}%

\begin{table*}[!ht]

\caption{\label{tab:lbrg}\normalsize{Mass accretion rate from Br$\gamma$ luminosity.}}
\begin{minipage}[c]{\textwidth}
    \renewcommand{\footnoterule}{}
    \renewcommand{\arraystretch}{1.7}
\centering
\begin{tabular}{lllllll}
\hline
\hline
HJD & $K_{\rm obs}$ & $EW({\rm Br}\gamma)_{\rm obs}$ & Reference & $EW({\rm Br}\gamma)_{\rm cs}$ & $L({\rm Br}\gamma)$ & $\log \dot{M}_{\rm acc}({\rm Br}\gamma)$ \\
 & (mag) & ($\AA$) & & ($\AA$) & ($10^{-3}$~L$_\odot$) & M$_\odot$~yr$^{-1}$  \\
\hline
23 Mar 2002 & $ 4.5^a$ & $-3.0$ & \citet{Brittain2007} & $ -7.2 \pm  1.2$ & $ 2.1 \pm  0.3$ & $-6.27 \pm 0.78$ \\ 
18 Jul 2002 & $ 4.5$ & $-3.1$ & \citet{Sitko2008} & $ -7.2 \pm  1.2$ & $ 2.1 \pm  0.3$ & $-6.26 \pm 0.78$ \\ 
09 Jun 2004 & $ 4.8$ & $-4.7$ & \citet{GarciaLopez2006} & $ -9.9 \pm  1.2$ & $ 2.3 \pm  0.3$ & $-6.23 \pm 0.76$ \\ 
06 Jul 2005 & $ 4.6$ & $-4.2$ & \citet{Sitko2008} & $ -8.8 \pm  1.2$ & $ 2.3 \pm  0.3$ & $-6.22 \pm 0.76$ \\ 
Mar 2008 & $ 4.8$ & $-4.3$ & \citet{Donehew2011} & $ -9.5 \pm  1.2$ & $ 2.2 \pm  0.3$ & $-6.24 \pm 0.76$ \\ 
13 May 2008 & $ 4.8$ & $-4.3$ & \citet{Donehew2011} & $ -9.5 \pm  1.2$ & $ 2.2 \pm  0.3$ & $-6.24 \pm 0.76$ \\ 
15 Jul 2009 & $ 4.8$ & $-3.2$ & \citet{Eisner2010} & $ -8.4 \pm  1.2$ & $ 2.0 \pm  0.3$ & $-6.29 \pm 0.78$ \\ 
12 Oct 2011 & $ 4.8$ & $-4.2$ & \citet{Mendigutia2013} & $ -9.4 \pm  1.2$ & $ 2.2 \pm  0.3$ & $-6.25 \pm 0.76$ \\ 
14 Oct 2011 & $ 4.5$ & $-3.3$ & \citet{Mendigutia2013} & $ -7.4 \pm  1.2$ & $ 2.2 \pm  0.4$ & $-6.25 \pm 0.77$ \\ 
16 Oct 2011 & $ 4.3$ & $-3.9$ & \citet{Mendigutia2013} & $ -7.3 \pm  1.2$ & $ 2.7 \pm  0.4$ & $-6.17 \pm 0.75$ \\ 
24 Mar 2012 & $ 4.3$ & $-3.7$ & \citet{Mendigutia2013} & $ -7.1 \pm  1.2$ & $ 2.6 \pm  0.4$ & $-6.18 \pm 0.75$ \\ 
17 May 2012 & $ 4.2$ & $-3.7$ & \citet{Mendigutia2013} & $ -6.7 \pm  1.2$ & $ 2.7 \pm  0.5$ & $-6.16 \pm 0.75$ \\ 
05 Jul 2012 & $ 4.7$ & $-4.3$ & This work & $ -9.2 \pm  1.2$ & $ 2.3 \pm  0.3$ & $-6.23 \pm 0.76$ \\ 
\hline

\hline
    \vspace{-20pt}
\end{tabular}
\end{minipage}

\end{table*}

\end{document}